\documentclass[12pt]{article}

\usepackage[a4paper,text={16.8cm,22.4cm}]{geometry}
\usepackage{amsmath,amsfonts,slashed,amssymb,tikz,bm,psfrag,graphicx,color,dsfont}
\usepackage{multicol}
\usepackage{float}

\RequirePackage[sort&compress,square,comma,numbers]{natbib}
\allowdisplaybreaks
\begin{document}

\begin{titlepage}

\begin{flushright}
\normalsize
September 28, 2018
\end{flushright}

\vspace{0.1cm}
\begin{center}
\Large\bf
QCD calculations of $B \to \pi, K$ form factors with higher-twist corrections
\end{center}

\vspace{0.5cm}
\begin{center}
{\bf Cai-Dian L\"{u}$^{a,b}$, Yue-Long Shen$^{c}$,
Yu-Ming Wang$^{d}$,  Yan-Bing Wei$^{d,a,b}$  } \\
\vspace{0.7cm}
{\sl   ${}^a$ \, Institute of High Energy Physics, CAS, P.O. Box 918(4) Beijing 100049,  China \\
 ${}^b$ \,  School of Physics, University of Chinese Academy of Sciences, Beijing 100049,  China \\
 ${}^c$ \, College of Information Science and Engineering,
Ocean University of China, Songling Road 238, Qingdao, 266100 Shandong, P.R. China \\
${}^d$ School of Physics, Nankai University, 300071 Tianjin, China \,
}
\end{center}

\vspace{0.2cm}
\begin{abstract}

We update QCD calculations of $B \to \pi, K$ form factors at large hadronic recoil by including the
subleading-power corrections from the higher-twist $B$-meson light-cone distribution amplitudes (LCDAs)
up to the twist-six accuracy and the strange-quark mass effects at leading-power in $\Lambda/m_b$
from the twist-two $B$-meson LCDA $\phi_B^{+}(\omega, \mu)$.  The higher-twist corrections from both the two-particle and
three-particle $B$-meson LCDAs are computed from the light-cone QCD sum rules (LCSR) at tree level.
In particular, we construct the local duality model for the twist-five and -six $B$-meson LCDAs,
in agreement  with the corresponding asymptotic behaviours at small quark and gluon momenta,
employing the QCD sum rules in heavy quark effective theory at leading order in $\alpha_s$.
The strange quark mass effects in semileptonic $B \to K$ form factors yield the leading-power contribution in the
heavy quark expansion, consistent with the power-counting analysis in
soft-collinear effective theory, and they are also computed from the LCSR approach due to the appearance of
the rapidity singularities. We demonstrate explicitly that the SU(3)-flavour symmetry breaking effects between
$B \to \pi$ and $B \to K$ form factors, free of the power suppression in $\Lambda/m_b$, are suppressed by
a factor of $\alpha_s(\sqrt{m_b \, \Lambda})$ in perturbative expansion, and they also respect the large-recoil
symmetry relations of the heavy-to-light form factors at least at one-loop accuracy.
An exploratory analysis of the obtained sum rules for $B \to \pi, K$ form factors
with two distinct models for the $B$-meson LCDAs indicates that the dominant higher-twist corrections
are from the Wandzura-Wilczek part of  the two-particle LCDA of twist five $g_B^{-}(\omega, \mu)$
instead of the three-particle $B$-meson LCDAs. The resulting SU(3)-flavour symmetry violation effects
of $B \to \pi, K$ form factors turn out to be insensitive to the non-perturbative models of $B$-meson LCDAs.
We further explore the phenomenological aspects of the semileptonic $B \to \pi \ell \nu$ decays and the rare
exclusive processes $B \to K \nu \nu$, including the determination of the CKM matrix element $|V_{ub}|$,
the normalized differential $q^2$ distributions and precision observables defined by the ratios of
branching fractions for the above-mentioned two channels in the same intervals of $q^2$.

\end{abstract}

\vfil

\end{titlepage}

\section{Introduction}
\label{sect:Intro}

Precision calculations of the semileptonic $B \to \pi, K$ form factors are of essential importance
for the determination of the CKM matrix element $|V_{ub}|$ exclusively and the theory description of
the flavour-changing neutral current process $B \to K \ell \ell$ in QCD.
At small  hadronic recoil the Lattice QCD calculations of these form factors have been performed in
\cite{Lattice:2015tia,Bailey:2015nbd,Bailey:2015dka},
using the gauge-field ensembles with (2+1)-flavour lattice configurations.
Diverse QCD techniques for computing the heavy-to-light form factors at large hadronic recoil
have been developed with distinct theory assumptions and approximations.
Factorization properties of exclusive  $B \to \pi, K$ form factors at large recoil have been extensively explored
in the framework of soft-collinear effective theory (SCET) \cite{Bauer:2000yr,Bauer:2001yt,Beneke:2002ph,Beneke:2002ni},
leading to the  QCD factorization formulae at leading power in the heavy quark expansion
\begin{eqnarray}
F_i^{B \to M}(E) &=& C_i(E) \, \xi_a(E) + \int d \tau \, C_i^{(B1)}(E, \tau) \, \Xi_a(\tau, E) \,, \qquad \\
\Xi_a(\tau, E)  &=&  {1 \over 4} \, \int_0^{\infty} \, d \omega \, \int_0^1 d v \, J_a(\tau; v, \omega) \,
\tilde{f}_B \, \phi_B^{+}(\omega) \, f_{M} \, \phi_{M}(v)\,.
\end{eqnarray}
The perturbative matching coefficients $C_i(E)$ and $C_i^{(B1)}(E, \tau)$ have been computed
at one loop \cite{Bauer:2000yr,Beneke:2004rc,Becher:2004kk},
and at two loops (only for $C_i(E)$) \cite{Bonciani:2008wf,Asatrian:2008uk,Beneke:2008ei,Bell:2008ws,Bell:2010mg}.
The jet functions $J_a$ from matching the $ \rm {SCET_{I}}$ matrix elements of the $B$-type operators onto
$ \rm {SCET_{II}}$ have been also determined at the one-loop accuracy \cite{Becher:2004kk,Hill:2004if,Beneke:2005gs}.
However, the soft-collinear factorization for the $ \rm {SCET_{I}}$ matrix elements $\xi_a(E)$ cannot be achieved
due to the emergence of end-point divergences, whose regularizations will introduce an unwanted connection between
the soft functions and the collinear functions (see \cite{Beneke:2003pa} for more discussions).
In this respect, the method of QCD sum rules on the light cone (LCSR) can be applied to evaluate the
non-perturbative form factors $\xi_a(E)$ directly by introducing the parton-hadron duality ansatz,
and the rapidity divergences appearing in the soft-collinear factorization are effectively regularized by the instinct
sum rule parameters. As a matter of fact, the heavy-to-light $B$-meson form factors $F_i^{B \to M}(E)$ themselves have
been widely investigated in the context of the LCSR approach with the light-meson light-cone distribution amplitudes (LCDAs)
\cite{Ball:2001fp,Ball:2004ye,Duplancic:2008ix} and
with the $B$-meson  LCDAs \cite{Khodjamirian:2005ea,Khodjamirian:2006st,DeFazio:2005dx,DeFazio:2007hw,Wang:2015vgv,Wang:2017jow}
(see \cite{Wang:2009hra,Feldmann:2011xf,Wang:2015ndk} for further extensions to the heavy-baryon decay form factors).
An alternative approach to investigate the heavy-to-light form factors is based upon
the transverse-momentum dependent (TMD) factorization \cite{Botts:1989kf,Li:1992nu}
with the assumption that the soft contribution does not contribute at leading power in $\Lambda/m_b$.
Technically, the rapidity divergences appearing in   SCET are regularized by  the intrinsic
momenta of the partons participating the hard reactions \cite{Li:2010nn,Li:2012nk,Li:2012md,Li:2013xna,He:2006ud,Lu:2009cm,Wang:2010ni,Li:2004ep}.
However, a rigorous proof of the TMD factorization for the hard exclusive processes is still not available
due to the absence of  a definite power counting scheme for  the intrinsic momenta \cite{Li:2014xda,Wang:2015qqr}.

Inspired by the experimental advances for precision measurements of the semileptonic $B \to \pi \ell \nu$ decays
as well as the electroweak penguin $B$-meson decays from Belle II \cite{Kou:2018nap}, we attempt to improve the theory predictions for
$B \to \pi, K$ form factors from the LCSR approach with the $B$-meson LCDAs  presented in \cite{DeFazio:2007hw,Wang:2015vgv},
where the next-to-leading-logarithmic (NLL) resummation improved sum rules for the leading-power contributions
were derived by applying QCD factorization for the corresponding vacuum-to-$B$-meson correlation function
and the dispersion relation technique. We summarize the main new ingredients of the present paper in the following.

\begin{itemize}

\item{We compute the subleading-power contributions to the semileptonic $B \to \pi, K$ form factors
from both the two-particle and three-particle higher-twist $B$-meson LCDAs with the LCSR method
at the twist-six accuracy. To this end, we adopt a complete parametrization of the three-particle
$B$-meson LCDAs proposed in \cite{Braun:2017liq}, which introduces eight independent invariant functions
in the light-cone limit (see \cite{Kawamura:2001jm} for the original incomplete parametrization).}

\item{We construct the local duality model for the twist-five and -six $B$-meson LCDAs $\Phi_5(\omega_1, \omega_2, \mu)$,
$\Psi_5(\omega_1, \omega_2, \mu)$, $\tilde{\Psi}_5(\omega_1, \omega_2, \mu)$ and $\Phi_6(\omega_1, \omega_2, \mu)$
employing the method of QCD sum rules at tree level. The local duality model for
the two-particle twist-five $B$-meson LCDA $\hat{g}_B^{-}(\omega, \mu)$ will be further derived with QCD equations of motion (EOM).
We verify explicitly that the obtained model for the higher-twist $B$-meson LCDAs is consistent with the corresponding asymptotic behaviours
at small quark and gluon momenta, which can be inferred from  the renormalization group (RG) equations of the corresponding
light-ray operators  at leading logarithmic (LL) accuracy.}

\item{We derive the leading-power contributions to $B \to K$ form factors from the strange-quark mass effects applying the
LCSR approach at next-to-leading-order (NLO) in $\alpha_s$. Our computation supports the early observation based upon the
power-counting analysis in SCET \cite{Leibovich:2003jd} that the SU(3)-flavour symmetry violation
between $B \to \pi$  and $B \to K$ form factors is not suppressed in the heavy quark limit
and the strange-quark mass effects will not give rise to the large-recoil symmetry breaking
of the heavy-to-light $B$-meson form factors.}

\end{itemize}

This paper is structured as follows.  We present QCD factorization formulae of the leading-twist
contributions to the vacuum-to-$B$-meson correlation function with an
interpolating current for the light pseudoscalar meson at one loop in
 Section \ref{sect: the leading-twist effect}, where the new jet function generated by the
 non-vanishing strange-quark mass is derived with the method of regions \cite{Beneke:1997zp}
 and the NLL resummation improved sum rules for $B \to \pi, K$ form factors are further obtained
 at leading-twist approximation.
A detailed calculation of the higher-twist contributions to the semileptonic $B$-meson form factors
from the LCSR method at tree level is presented in Section \ref{sect: the higher-twist effect},
where the power counting of both the two-particle and three-particle subleading-power contributions
is further  discussed.  Inspecting the correlation functions of the light-ray operators
defining the higher-twist $B$-meson LCDAs and suitable local currents in heavy-quark effective theory (HQET),
we derive the QCD sum rules for the twist-five and -six $B$-meson LCDAs at leading-order (LO) in $\alpha_s$
in Section \ref{sect: QCDSR for higher-twist DAs}, where the local duality model for these distribution
amplitudes is obtained by taking the limit $M^2 \to \infty$. We explore the phenomenological implications
of the new sum rules for  $B \to \pi, K$ form factors with distinct models of the $B$-meson LCDAs
in Section \ref{sect: numerical analysis}, including the numerical impacts of the higher-twist corrections,
the model dependence of the SU(3)-flavour symmetry breaking effects,
a comparison of the large-recoil symmetry violation with that predicted
from the QCD factorization approach \cite{Beneke:2000wa}, the determination of the CKM matrix element $|V_{ub}|$,
and the differential $q^2$ distributions of $B \to \pi \ell \nu$ and $B \to K \nu \nu$.
We will conclude in Section \ref{sect: summary} with a summary of our main observations and perspectives
on the future developments.

\section{The leading-twist contributions to the LCSR  at ${\cal O}(\alpha_s)$}
\label{sect: the leading-twist effect}

Following the procedure presented in \cite{Wang:2015vgv}, the sum rules for
 $B \to \pi, K$ form factors can be constructed from the following
 vacuum-to-$B$-meson correlation function
 \begin{eqnarray}
\Pi_{\mu}(n \cdot p,\bar n \cdot p)&=& \int d^4x ~e^{i p\cdot x}
\langle 0 |T\left\{\bar{d}(x) \not n \, \gamma_5 \, q(x), \,\,
\bar{q}(0) \, \Gamma_\mu  \, b(0) \right\}|\bar B(p+q) \rangle \nonumber \\
&=&   \left\{
\begin{array}{l}
\Pi(n \cdot p,\bar n \cdot p) \,  n_\mu +\widetilde{\Pi}(n \cdot p,\bar n \cdot p) \, \bar n_\mu \,, \qquad
\Gamma_{\mu}  =  \gamma_{\mu} \vspace{0.4 cm} \\
\Pi_T(n \cdot p,\bar n \cdot p) \, \left [n_{\mu} - \frac{n \cdot q}  {m_B} \, \bar n_{\mu} \right ] \,,
 \qquad  \hspace{1.0 cm} \Gamma_{\mu}  =  \sigma_{\mu \nu} \, q^{\mu}
\end{array}
 \hspace{0.5 cm} \right.
\,
\label{correlator: definition}
\end{eqnarray}
for the two different $b \to q$ weak currents in QCD, where the light pseudoscalar meson is interpolated by
an axial-vector current carrying the four-momentum $p$ and $p+q \equiv m_B \, v$ indicates the four-momentum of the $B$ meson.
We further introduce two light-cone vectors $n_{\mu}$ and $\bar n_{\mu}$ satisfying $n^2=\bar n^2=0$ and $n \cdot \bar n=2$,
and employ the following power counting scheme
\begin{eqnarray}
 n \cdot p \sim {\cal O}(m_B) \,, \qquad  \bar n \cdot p \sim m_s \sim {\cal O}(\Lambda) \,.
\end{eqnarray}
Applying the method of regions one can establish QCD factorization formulae for the correlation function
(\ref{correlator: definition}) at leading power in $\Lambda/m_b$
\begin{eqnarray}
\Pi &=& \tilde{f}_B(\mu) \, m_B \sum \limits_{k=\pm} \,
C^{(k)}(n \cdot p, \mu) \, \int_0^{\infty} {d \omega \over \omega- \bar n \cdot p}~
J^{(k)}\left({\mu^2 \over n \cdot p \, \omega},{\omega \over \bar n \cdot p}\right) \,
\phi_B^{k}(\omega,\mu)  \,,
\nonumber \\
\widetilde{\Pi} &=& \tilde{f}_B(\mu) \, m_B \sum \limits_{k=\pm} \,
\widetilde{C}^{(k)}(n \cdot p, \mu) \, \int_0^{\infty} {d \omega \over \omega- \bar n \cdot p}~
\widetilde{J}^{(k)}\left({\mu^2 \over n \cdot p \, \omega},{\omega \over \bar n \cdot p}\right) \,
\phi_B^{k}(\omega,\mu)  \,,
\nonumber \\
\Pi_T &=& - {i \over 2} \,  \tilde{f}_B(\mu) \, m_B^2 \,  \sum \limits_{k=\pm} \,
C^{(k)}_T(n \cdot p, \mu, \nu) \, \int_0^{\infty} {d \omega \over \omega- \bar n \cdot p}~
J^{(k)}_T \left({\mu^2 \over n \cdot p \, \omega},{\omega \over \bar n \cdot p}\right) \,
\phi_B^{k}(\omega,\mu) \,.
\label{QCD factorization formula of correlator at LT}
\end{eqnarray}
The $B$-meson LCDAs in coordinate space are defined by the renormalized matrix element
of the following light-cone operator in HQET \cite{Grozin:1996pq}
\begin{eqnarray}
&& \langle  0 | \left (\bar d \, Y_s \right)_{\beta} (\tau \, \bar{n}) \,
\left (Y_s^{\dag} \, h_v \right )_{\alpha}(0)| \bar B(v)\rangle \nonumber \\
&& = - \frac{i \tilde f_B(\mu) \, m_B}{4}  \bigg \{ \frac{1+ \! \not v}{2} \,
\left [ 2 \, \tilde{\phi}_{B}^{+}(\tau, \mu) + \left ( \tilde{\phi}_{B}^{-}(\tau, \mu)
-\tilde{\phi}_{B}^{+}(\tau, \mu)  \right )  \! \not n \right ] \, \gamma_5 \bigg \}_{\alpha \beta}\,,
\label{def: two-particle B-meson DAs}
\end{eqnarray}
where the soft Wilson line is given by
\begin{eqnarray}
Y_s(\tau \, \bar n)= {\rm P} \, \left \{ {\rm  Exp} \left [   i \, g_s \,
\int_{- \infty}^{\tau} \, dx \,  \bar n  \cdot A_{s}(x \, \bar n) \right ]  \right \} \,.
\label{def: soft gauge link}
\end{eqnarray}
The renormalization-scale dependent HQET decay constant $\tilde f_B(\mu)$ can be expressed
in terms of the QCD decay constant $f_B$
\begin{eqnarray}
\tilde{f}_B(\mu)= \left \{  1 -  {\alpha_s(\mu) \, C_F \over 4 \, \pi} \,
\left [3\, \ln{\mu \over m_b} + 2  \right ] \right \}^{-1} \, f_B \,.
\label{HQET matching of fB}
\end{eqnarray}
We can further determine the renormalized hard functions and jet functions entering
the factorization formulae (\ref{QCD factorization formula of correlator at LT}) at the one-loop accuracy
\begin{eqnarray}
C^{(+)} &=&  \widetilde{C}^{(+)} = C_T^{(+)} = 1 \,, \qquad
C^{(-)} =  {\alpha_s \, C_F \over 4 \pi} \, {1 \over \bar r} \,
\left [ 1 + { r \over \bar r} \, \ln r \right ] \,, \nonumber \\
\widetilde{C}^{(-)} &=&  1-  {\alpha_s \, C_F \over 4 \pi} \,
\left [ 2 \, \ln^2 {\mu \over n \cdot p} + 5 \, \ln {\mu \over m_b}
- \ln^2 r - 2 \, {\rm Li}_2 \left (- { \bar r \over r } \right )
+ { 2 - r \over r-1}  \, \ln r + {\pi^2 \over 12} + 5 \right ] \,, \nonumber \\
C_T^{(-)} &=&   1 +  {\alpha_s \, C_F \over 4 \pi} \,
\left [ -2 \, \ln {\nu \over m_b} - 2 \, \ln^2 {\mu \over n \cdot p}
- 5 \, \ln {\mu \over n \cdot p}  - 2 \, {\rm Li}_2 (1-r)
- {3 - r \over 1 -r} \, \ln r -  {\pi^2 \over 12} - 6 \right ] \,, \nonumber \\
J^{(+)} &=&   {\alpha_s \, C_F \over 4 \pi} \, \left (1 -  {\bar n \cdot p \over \omega} \right ) \,
\ln \left (1- {\omega \over \bar n \cdot p} \right ) \,, \nonumber \\
\widetilde{J}^{(+)} &=&   {\alpha_s \, C_F \over 4 \pi} \,
\left [ r \, \left (1 -  {\bar n \cdot p \over \omega} \right ) + {m_q \over \omega} \right ] \,
\ln \left (1- {\omega \over \bar n \cdot p} \right ) \,, \nonumber  \\
J_T^{(+)} &=& {\alpha_s \, C_F \over 4 \pi} \,
\left [ - \, \left (1 -  {\bar n \cdot p \over \omega} \right ) + {m_q \over \omega} \right ] \,
\ln \left (1- {\omega \over \bar n \cdot p} \right )  \,, \nonumber  \\
J^{(-)} &=&  1  \,, \nonumber  \\
\widetilde{J}^{(-)} &=&   J_T^{(-)}
=  1 + \frac{\alpha_s \, C_F}{4 \, \pi} \,
\bigg [ \ln^2 { \mu^2 \over  n \cdot p (\omega- \bar n \cdot p) }
- 2 \ln {\bar n \cdot p -\omega \over \bar n \cdot p } \, \ln { \mu^2 \over  n \cdot p (\omega- \bar n \cdot p) }
\,  \nonumber \\
&& - \ln^2 {\bar n \cdot p -\omega \over \bar n \cdot p }
- \left ( 1 +  {2 \bar n \cdot p \over \omega} \right )  \ln {\bar n \cdot p -\omega \over \bar n \cdot p }
-{\pi^2 \over 6} -1 \bigg ] \,,
\label{matching coefficients of the LP contributions}
\end{eqnarray}
where $\nu$ refers to the renormalization scale of the QCD tensor current and
we have also introduced the conventions
\begin{eqnarray}
r= n \cdot p / m_b \,, \qquad  \bar r =1-r.
\end{eqnarray}

Several remarks on the resulting perturbative matching coefficients are in order.

\begin{itemize}

\item{The hard matching coefficients appearing in the QCD factorization formulae for the
vacuum-to-$B$-meson correlation function (\ref{QCD factorization formula of correlator at LT})
are apparently identical to the short-distance functions from representing the
corresponding QCD weak currents in ${\rm SCET_{I}}$.
Employing the perturbative matching  for heavy-to-light currents
displayed in \cite{Bauer:2000yr}
\begin{eqnarray}
\bar q \, \gamma_{\mu} \, b  & \to & \left [C_4 \, \bar n_{\mu} + C_5 \, v_{\mu} \right ] \,
\bar \xi_{\bar n} \, W_{hc} \,\, Y_s^{\dag} \, h_v + ...,  \nonumber \\
\bar q \, \sigma_{\mu \nu} \, q^{\nu}\, b  & \to &  (-i) \, C_{11} \,
(v_{\mu} \bar n_{\nu} - v_{\nu} \bar n_{\mu}) \, q^{\nu}  \,
\bar \xi_{\bar n} \, W_{hc} \,\, Y_s^{\dag} \, h_v + ... \,,
\end{eqnarray}
 one can readily determine the following relations for the hard functions
\begin{eqnarray}
C^{(-)}= {1 \over 2} \, C_5 \,,   \qquad    \widetilde{C}^{(-)}= C_4 + {1 \over 2} \, C_5  \,,
\qquad C_T^{(-)}=C_{11} \,,
\end{eqnarray}
which can be further verified by comparing the explicit expressions of $C_4$, $C_5$ and $C_{11}$
obtained in  \cite{Bauer:2000yr}
with the results presented in (\ref{matching coefficients of the LP contributions}).
}

\item{The nonvanishing light-quark mass gives rise to the leading-power contribution to the jet functions
in the heavy quark expansion, which is independent of the Dirac structures of the QCD weak currents.
Our calculation supports the power counting analysis for the light-quark mass effects in semileptonic
$B$-meson decay form factors  at large recoil in the framework of SCET \cite{Leibovich:2003jd}.
Inspecting the diagrammatical representation of the two-particle contributions to
correlation function (\ref{correlator: definition}) at NLO in QCD, we observe that the above-mentioned
SU(3)-flavour symmetry breaking effect solely comes  from  the QCD correction to the
light-pseudoscalar-meson vertex diagram (see figure 2(a) of \cite{Wang:2015vgv}).
It immediately follows that the light-quark-mass dependent jet function entering the factorization formulae
(\ref{QCD factorization formula of correlator at LT}) is universal for the vacuum-to-$B$-meson correlation functions
with different weak currents.}

\end{itemize}

We will proceed to perform the summation of parametrically large logarithms appearing in the factorization
formulae  (\ref{QCD factorization formula of correlator at LT}) employing the RG equations in momentum space.
Taking the factorization scale $\mu$ as a hard-collinear scale $\mu_{hc} \sim \sqrt{\Lambda \, m_b}$
and solving the evolution equations at NLL accuracy leads to
\begin{eqnarray}
\widetilde{C}^{(-)}(n \cdot p, \mu) &=& U_1(n \cdot p, \mu_{h1}, \mu) \,\, \widetilde{C}^{(-)}(n \cdot p, \mu_{h1})\,, \nonumber \\
C^{(-)}_{T}(n \cdot p, \mu, \nu)  &=& U_1(n \cdot p, \mu_{h1}, \mu) \,\, U_3(\nu_{h}, \nu) \,\,
C^{(-)}_{T}(n \cdot p, \mu_{h1}, \nu_{h}) \,,  \nonumber \\
\tilde{f}_B(\mu)&=& \, U_2(\mu_{h2}, \mu) \, \tilde{f}_B(\mu_{h2}) \,\,,
\end{eqnarray}
where the explicit expressions of the evolution functions $U_1$ and $U_2$ can be found in \cite{Beneke:2011nf,Wang:2016qii}
and the QCD evolution factor $U_3(\nu_{h}, \nu)$ is given by
\begin{eqnarray}
U_3(\nu_{h}, \nu) &=& {\rm Exp}  \bigg [ \int_{\alpha_s(\nu_{h})}^{\alpha_s(\nu)} \,
d \alpha_s \, \frac{\gamma_T(\alpha_s)}{\beta(\alpha_s)} \bigg ] \, \nonumber \\
&=& z^{- \frac{\gamma_T^{(0)}}{2 \,\beta_0}} \bigg [1+ \frac{\alpha_s(\nu_{h})}{4 \pi}  \,
\left (  {\gamma_T^{(1)} \over 2 \, \beta_0} - {\gamma_T^{(0)} \, \beta_1 \over 2 \, \beta_0^2 } \right ) (1-z)
+{\cal O}(\alpha_s^2) \bigg ]\,,
\end{eqnarray}
with $z=\alpha_s(\nu)/\alpha_s(\nu_h)$. The anomalous dimension $\gamma_T(\alpha_s)$ for the tensor current
at the two-loop accuracy is \cite{Bell:2010mg}
\begin{eqnarray}
\gamma_T(\alpha_s) &=& \sum_{n=0}^{\infty} \, \left ( {\alpha_s(\mu) \over 4 \, \pi} \right )^{n+1} \,
\gamma_T^{(n)} \,, \qquad \gamma_T^{(0)} = - 2\, C_F  \,, \nonumber \\
\gamma_T^{(1)} &=&   C_F \, \left [ 19 \, C_F \,  -{257 \over 9} \, C_A
+  {52 \over 9} \, n_f \, T_F \right ]\,.
\end{eqnarray}
Since the hard-collinear scale $\mu_{hc}$ is quite close to the soft scale $\mu_0$
of the $B$-meson LCDAs numerically and the two-loop evolution equations of the two-particle
$B$-meson distribution amplitudes $\phi_B^{\pm}(\omega, \mu)$  are not available yet,
we will not perform the  NLL resummmation for the logarithms of $\mu/\mu_0$ due to the RG evolution
of $\phi_B^{\pm}(\omega, \mu)$ (see also \cite{Beneke:2011nf}).
It is then straightforward to write down the (partial) NLL resummation improved QCD factorization formulae
for the correlation function (\ref{correlator: definition})
\begin{eqnarray}
\Pi &=&  \left [ U_2(\mu_{h2}, \mu) \, \tilde{f}_B(\mu_{h2}) \right ]\, m_B  \,
\, \bigg \{  \int_0^{\infty} {d \omega \over \omega- \bar n \cdot p}~
J^{(+)}\left({\mu^2 \over n \cdot p \, \omega},{\omega \over \bar n \cdot p}\right) \,
\phi_B^{+}(\omega,\mu)  \nonumber \\
&&  +  \, C^{(-)}(n \cdot p, \mu)  \,
\int_0^{\infty} {d \omega \over \omega- \bar n \cdot p}~ \,
\phi_B^{-}(\omega,\mu)  \,  \bigg \}\,,
\nonumber  \\
\widetilde{\Pi} &=&  \left [ U_2(\mu_{h2}, \mu) \, \tilde{f}_B(\mu_{h2}) \right ]\, m_B  \,
\, \bigg \{  \int_0^{\infty} {d \omega \over \omega- \bar n \cdot p}~
\widetilde{J}^{(+)}\left({\mu^2 \over n \cdot p \, \omega},{\omega \over \bar n \cdot p}\right) \,
\phi_B^{+}(\omega,\mu)  \nonumber \\
&&  +  \,\left [  U_1(n \cdot p, \mu_{h1}, \mu) \,\, \widetilde{C}^{(-)}(n \cdot p, \mu_{h1}) \right ] \,
\int_0^{\infty} {d \omega \over \omega- \bar n \cdot p}~
\widetilde{J}^{(-)}\left({\mu^2 \over n \cdot p \, \omega},{\omega \over \bar n \cdot p}\right) \,
\phi_B^{-}(\omega,\mu)  \,  \bigg \}\,,
\nonumber  \\
\Pi_T &=& - {i \over 2} \, \left [ U_2(\mu_{h2}, \mu) \, \tilde{f}_B(\mu_{h2}) \right ]\, m_B^2  \,
\, \bigg \{  \int_0^{\infty} {d \omega \over \omega- \bar n \cdot p}~
J_T^{(+)}\left({\mu^2 \over n \cdot p \, \omega},{\omega \over \bar n \cdot p}\right) \,
\phi_B^{+}(\omega,\mu)  \nonumber \\
&&  +  \,\left [ U_1(n \cdot p, \mu_{h1}, \mu) \,\, U_3(\nu_{h}, \nu) \,\,
C^{(-)}_{T}(n \cdot p, \mu_{h1}, \nu_{h})  \right ] \, \nonumber  \\
&& \hspace{0.5 cm} \times \, \int_0^{\infty} {d \omega \over \omega- \bar n \cdot p}~
J_T^{(-)}\left({\mu^2 \over n \cdot p \, \omega},{\omega \over \bar n \cdot p}\right) \,
\phi_B^{-}(\omega,\mu)  \,  \bigg \} \,.
\label{Resummation improved factorization formula of correlator at LT}
\end{eqnarray}

Employing the standard definitions for the $B \to P$ form factors (with $P=\pi, \,\, K$)
and the decay constant of the pseudoscalar meson
\begin{eqnarray}
\langle P(p)|  \bar q \, \gamma_{\mu} \, b| \bar B (p+q)\rangle
&=& f_{B \to P}^{+}(q^2) \, \left [ 2 p + q -\frac{m_B^2-m_{P}^2}{q^2} q  \right ]_{\mu}
+  f_{B \to P}^{0}(q^2) \, \frac{m_B^2-m_{P}^2}{q^2} q_{\mu} \,, \nonumber \\
\langle P(p)|  \bar q \, \sigma_{\mu \nu} \, q^{\nu}\, b| \bar B (p+q)\rangle
&=& i \, {f_{B \to P}^{T}(q^2) \over m_B + m_P}\, \left [ q^2   \,\, (2 p + q)_{\mu} \,
-(m_B^2-m_{P}^2) \, q_{\mu}  \right ] \,, \nonumber \\
\langle 0 |\bar d \! \not n \, \gamma_5 \, q |  P(p)  \rangle &=&  i \, n \cdot p \, f_{P} \,,
\end{eqnarray}
we can readily derive the hadronic representations of the correlation function (\ref{correlator: definition})
\begin{eqnarray}
\Pi_{\mu, V}(n \cdot p,\bar n \cdot p) &=& \frac{f_{P} \, m_B}{2 \, (m_{P}^2/ n \cdot p - \bar n \cdot p)}
\bigg \{  \bar n_{\mu} \, \left [ \frac{n \cdot p}{m_B} \, f_{B \to P}^{+} (q^2) + f_{B \to P}^{0} (q^2)  \right ]
\nonumber \\
&& \hspace{0.4 cm} +  \,  n_{\mu} \, \frac{m_B}{n \cdot p-m_B}  \, \,
\left [ \frac{n \cdot p}{m_B} \, f_{B \to P}^{+} (q^2) -  f_{B \to P}^{0} (q^2)  \right ] \bigg \} \, \nonumber \\
&& \hspace{0.4 cm} + \int_{\omega_s}^{+\infty}   \, \frac{d \omega^{\prime} }{\omega^{\prime} - \bar n \cdot p - i 0} \,
\left [ \rho_{V, 1}^{h}(\omega^{\prime}, n \cdot p)  \, n_{\mu} \,
+\rho_{V, 2}^{h}(\omega^{\prime}, n \cdot p)  \, \bar{n}_{\mu}  \right ] \,, \nonumber \\
\Pi_{\mu, T}(n \cdot p,\bar n \cdot p) &=& - i \, \frac{f_{P} \, n \cdot p }{2 \, (m_{P}^2/ n \cdot p - \bar n \cdot p)}
\, {m_B^2 \over m_B+m_P} \, \left [ n_{\mu} -  {n \cdot q \over m_B} \, \bar n_{\mu} \right ]  \, f_{B \to P}^{T} (q^2) \nonumber \\
&& \hspace{0.4 cm} + \int_{\omega_s}^{+\infty}   \, \frac{d \omega^{\prime} }{\omega^{\prime} - \bar n \cdot p - i 0} \,
\, \left [ n_{\mu} -  {n \cdot q \over m_B} \, \bar n_{\mu} \right ]  \,
\rho_{T}^{h}(\omega^{\prime}, n \cdot p) \,,
\label{hadronic representations}
\end{eqnarray}
where $\Pi_{\mu, V}$ and $\Pi_{\mu, T}$ correspond to $\Gamma_{\mu}=\gamma_{\mu}$
and $\Gamma_{\mu}=\sigma_{\mu \nu} \, q^{\nu}$
for the Dirac structure of the weak current $\bar{q}(0) \, \Gamma_\mu  \, b(0) $
in the definition (\ref{correlator: definition}), respectively.
Matching the hadronic dispersion relations (\ref{hadronic representations})
and the resummation improved factorization formulae
(\ref{Resummation improved factorization formula of correlator at LT}) with the aid of the
parton-hadron duality ansatz and implementing the Borel transformation with respect
to the variable $\bar n \cdot p \to \omega_M$ gives rise to the NLL LCSR for
$B \to P$ form factors at leading power in the heavy quark expansion
\begin{eqnarray}
&& f_{P} \,\, {\rm exp} \left [- {m_{P}^2 \over n \cdot p \,\, \omega_M} \right ] \,\,
\left \{ \frac{n \cdot p} {m_B} \, f_{B \to P}^{+, \, \rm 2PNLL}(q^2)
\,, \,\,\,   f_{B \to P}^{0, \, \rm 2PNLL}(q^2)  \right \}  \,  \nonumber \\
&& =   \left [ U_2(\mu_{h2}, \mu) \, \tilde{f}_B(\mu_{h2}) \right ]
\,\, \int_0^{\omega_s} \,\, d \omega^{\prime} \, e^{-\omega^{\prime}/\omega_M} \,  \nonumber \\
&&  \hspace{0.4 cm} \times \bigg \{\widetilde{\phi}_{B, \, \rm {eff}}^{+} (\omega^{\prime}, \mu)
+  \, \left [  U_1(n \cdot p, \mu_{h1}, \mu) \,\, \widetilde{C}^{(-)}(n \cdot p, \mu_{h1}) \right ]\,
\widetilde{ \phi}_{B, \, \rm {eff}}^{-} (\omega^{\prime}, \mu) \nonumber \\
&& \hspace{0.8 cm} \pm \, { n \cdot p - m_B \over m_B} \,
\left [\phi_{B, \, \rm {eff}}^{+} (\omega^{\prime}, \mu)
+ C^{(-)}(n \cdot p, \mu_{h1}) \, \phi_{B, \, \rm {eff}}^{-} (\omega^{\prime}, \mu)   \right ]  \bigg \} \,, \nonumber \\
&& f_{P} \,\,  {\rm exp} \left [- {m_{P}^2 \over n \cdot p \,\, \omega_M} \right ]  \,\,
\frac{n \cdot p} {m_B+m_P} \,  f_{B \to P}^{T, \, \rm 2PNLL}(q^2) \,  \nonumber \\
&& =   \left [ U_2(\mu_{h2}, \mu) \, \tilde{f}_B(\mu_{h2}) \right ]
\,\, \int_0^{\omega_s} \,\, d \omega^{\prime} \, e^{-\omega^{\prime}/\omega_M} \,  \nonumber \\
&&  \hspace{0.4 cm} \times \bigg \{ \widehat{\phi}_{B, \, \rm {eff}}^{+} (\omega^{\prime}, \mu)
+  \,\left [ U_1(n \cdot p, \mu_{h1}, \mu) \,\, U_3(\nu_{h}, \nu) \,\,
C^{(-)}_{T}(n \cdot p, \mu_{h1}, \nu_{h})  \right ] \,
\widetilde{ \phi}_{B, \, \rm {eff}}^{-} (\omega^{\prime}, \mu)  \bigg \} \,,
\label{NLO sum rules of B to P form factors}
\end{eqnarray}
where we have defined the effective $B$-meson ``distribution amplitudes" for brevity
\begin{eqnarray}
\widetilde{\phi}_{B, \, \rm {eff}}^{+} (\omega^{\prime}, \mu) &=&
{\alpha_s \, C_F \over 4 \, \pi} \,
\left [ r \, \int_{\omega^{\prime}}^{\infty} \, d \omega \,
{\phi_B^{+}(\omega, \mu) \over \omega}
- m_q \, \int_{\omega^{\prime}}^{\infty} \, d \omega \,
\ln \left ( {\omega -  \omega^{\prime} \over \omega^{\prime}} \right )   \,\,
{d \over d \omega} \, {\phi_B^{+}(\omega, \mu) \over \omega}  \right ] \,,  \nonumber \\
\widetilde{\phi}_{B, \, \rm {eff}}^{-} (\omega^{\prime}, \mu) &=&
\phi_{B}^{-}(\omega^{\prime}, \mu)
+  \frac{\alpha_s \, C_F}{4 \, \pi} \,\, \bigg \{ \int_0^{\omega^{\prime}} \,\, d \omega \,\,\,
\left [ {2 \over \omega - \omega^{\prime}}  \,\,\, \left (\ln {\mu^2 \over n \cdot p \, \omega^{\prime}}
- 2 \, \ln {\omega^{\prime} - \omega \over \omega^{\prime}} \right )\right ]_{\oplus} \,
\phi_{B}^{-}(\omega, \mu)  \nonumber \\
&& - \int_{\omega^{\prime}}^{\infty} \,\, d \omega \,\,\,
\bigg [ \ln^2 {\mu^2 \over n \cdot p \, \omega^{\prime}}
- \left ( 2 \, \ln {\mu^2 \over n \cdot p \, \omega^{\prime}}  + 3 \right ) \,\,
\ln {\omega - \omega^{\prime} \over \omega^{\prime}}
+ \, 2 \,\, \ln {\omega \over \omega^{\prime}}   + {\pi^2 \over 6} - 1 \bigg ]
\nonumber \\
&& \hspace{0.5 cm}  \times \, {d \phi_{B}^{-}(\omega, \mu) \over d \omega}  \bigg \}  \,,   \nonumber \\
\phi_{B, \, \rm {eff}}^{+} (\omega^{\prime}, \mu) &=&  {\alpha_s \, C_F \over 4 \, \pi} \,
\int_{\omega^{\prime}}^{\infty} \, d \omega \, {\phi_B^{+}(\omega, \mu) \over \omega} \,,
\qquad
\phi_{B, \, \rm {eff}}^{-} (\omega^{\prime}, \mu) = \phi_B^{-}(\omega^{\prime}, \mu) \,, \nonumber \\
\widehat{\phi}_{B, \, \rm {eff}}^{+} (\omega^{\prime}, \mu) &=&
{\alpha_s \, C_F \over 4 \, \pi} \,
\left [ - \int_{\omega^{\prime}}^{\infty} \, d \omega \,
{\phi_B^{+}(\omega, \mu) \over \omega}
- m_q \, \int_{\omega^{\prime}}^{\infty} \, d \omega \,
\ln \left ( {\omega -  \omega^{\prime} \over \omega^{\prime}} \right )   \,\,
{d \over d \omega} \, {\phi_B^{+}(\omega, \mu) \over \omega}  \right ] \,.
\label{effective B-meson DAs}
\end{eqnarray}
The plus function entering (\ref{effective B-meson DAs}) is further given by
\begin{eqnarray}
\int_0^{\infty} \, d \omega \, \left [ f(\omega, \omega^{\prime}) \right ]_{\oplus} \, g(\omega)
= \int_0^{\infty} \, d \omega \,  f(\omega, \omega^{\prime})
\left [ g(\omega) - g(\omega^{\prime}) \right ]   \,.
\end{eqnarray}
It is evident that the light-quark-mass dependent effects respect the large-recoil symmetry relations
for the soft contribution to  the heavy-to-light form factors in the absence of corrections of order
$\alpha_s$ and $\Lambda/m_b$.

\section{The higher-twist contributions to the LCSR}
\label{sect: the higher-twist effect}

\begin{figure}
\begin{center}
\includegraphics[width=0.35 \columnwidth]{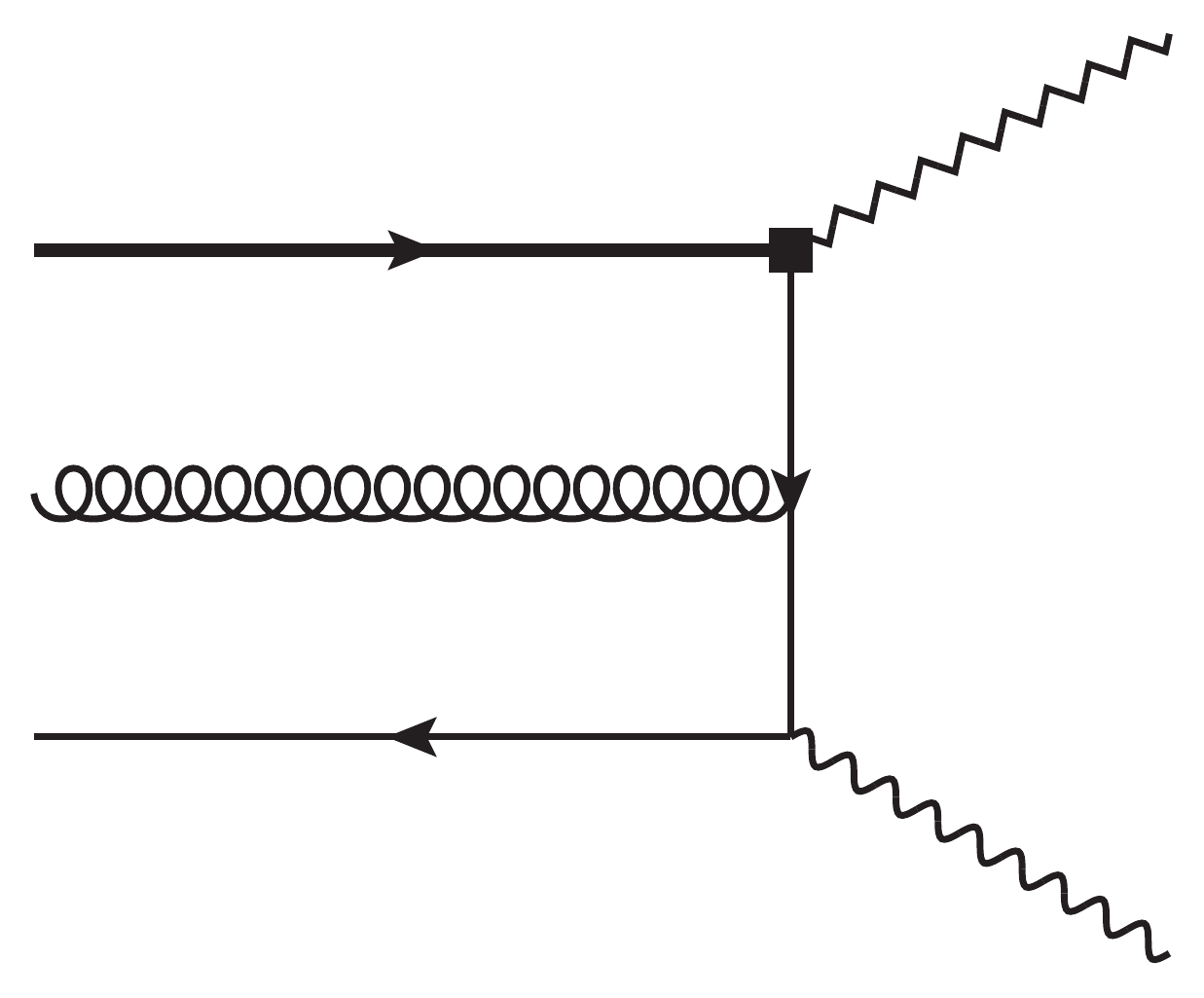}
\vspace*{0.1cm}
\caption{Diagrammatical representation of the three-particle higher-twist corrections
to the vacuum-to-$B$-meson correlation function (\ref{correlator: definition}).
The square box indicates the insertion of the weak vertex $\bar{q} \, \Gamma_\mu  \, b$,
and the waveline represents the interpolating current
$\bar{d} \not \! \! n \, \gamma_5 \, q$ for the light-pseudoscalar meson.  }
\label{fig: correlator for 3P at LO}
\end{center}
\end{figure}

We turn to compute the higher-twist corrections to $B \to \pi, K$ form factors from
both the two-particle and three-particle $B$-meson LCDAs employing a complete parametrization
of the corresponding three-particle light-cone matrix element and the EOM constraints of
the higher-twist LCDAs presented in  \cite{Braun:2017liq}.
To achieve this goal, we make use of the light-cone expansion of the quark propagator
in the background gluon field \cite{Balitsky:1987bk}
\begin{eqnarray}
\langle 0 | {\rm T} \, \{\bar q (x), q(0) \} | 0\rangle
 \supset   i \, g_s \, \int_0^{\infty} \,\, {d^4 k \over (2 \pi)^4} \, e^{- i \, k \cdot x} \,
\int_0^1 \, d u \, \left  [ {u \, x_{\mu} \, \gamma_{\nu} \over k^2 - m_q^2}
 - \frac{(\not \! k + m_q) \, \sigma_{\mu \nu}}{2 \, (k^2 - m_q^2)^2}  \right ]
\, G^{\mu \nu}(u \, x) \,, \hspace{0.4 cm}
\end{eqnarray}
where we only keep the one-gluon part without the covariant derivative of the $G_{\mu \nu}$  terms.
Evaluating the tree-level diagram displayed in figure \ref{fig: correlator for 3P at LO},
it is  straightforward to derive the three-particle higher-twist corrections to the vacuum-to-$B$-meson
correlation function (\ref{correlator: definition})

\begin{eqnarray}
\Pi_{\mu, V}^{(3P)}(n \cdot p,\bar n \cdot p) &=&
-{\tilde{f}_B(\mu) \, m_B \over n \cdot p } \, \int_0^{\infty} \, d \omega_1  \, \int_0^{\infty} \, d \omega_2 \,
\int_0^1 d u \, {1 \over \left [\bar n \cdot p - \omega_1 - u \, \omega_2 \right ]^2} \nonumber \\
&&  \times \, \bigg \{ \bar n_{\mu} \, \left [ \rho_{\bar n, \rm{LP}}^{(3P)}(u, \omega_1, \omega_2, \mu)
+ {m_q \over n \cdot p} \, \rho_{\bar n, \rm{NLP}}^{(3P)}(u, \omega_1, \omega_2, \mu) \right ] \nonumber \\
&& +\,  n_{\mu} \, \left [ \rho_{n, \rm{LP}}^{(3P)}(u, \omega_1, \omega_2, \mu)
+ {m_q \over n \cdot p} \, \rho_{n, \rm{NLP}}^{(3P)}(u, \omega_1, \omega_2, \mu) \right ] \bigg \}   \,, \nonumber \\
\Pi_{\mu, T}^{(3P)}(n \cdot p,\bar n \cdot p) &=&  {i \over 2} \, {\tilde{f}_B(\mu) \, m_B^2 \over  n \cdot p } \,
\, \left [ n_{\mu} -  {n \cdot q \over m_B} \, \bar n_{\mu} \right ]
\, \int_0^{\infty} \, d \omega_1  \, \int_0^{\infty} \, d \omega_2 \,
\int_0^1 d u \, {1 \over \left [\bar n \cdot p - \omega_1 - u \, \omega_2 \right ]^2} \nonumber \\
&& \times  \bigg \{  \rho_{T, \rm{LP}}^{(3P)}(u, \omega_1, \omega_2, \mu)
+ {m_q \over n \cdot p} \, \rho_{T, \rm{NLP}}^{(3P)}(u, \omega_1, \omega_2, \mu)  \bigg \}  \,,
\end{eqnarray}
where we have taken into account the SU(3)-flavour symmetry breaking effect due to the light-quark mass.
In contrast to the two-particle contributions to the correlation function (\ref{correlator: definition}),
the light-quark mass dependent terms of the three-particle corrections at LO in $\alpha_s$
are suppressed by one power of $\Lambda/m_b$.
The explicit expressions of $\rho_{i, \rm{LP}}^{(3P)}$ and $\rho_{i, \rm{NLP}}^{(3P)}$
($i=n\,, \bar n \,, T$) are given by
\begin{eqnarray}
\rho_{\bar n, \rm{LP}}^{(3P)} &=& (1-2 \, u) \, \left [ X_A - \Psi_A- 2 \, Y_A\right ]
-\tilde{X}_A  - \Psi_V  + 2\, \tilde{Y}_A  \,, \nonumber \\
\rho_{\bar n, \rm{NLP}}^{(3P)} &=& 2 \, \left [ \Psi_A - \Psi_V \right ]
+ 4 \, \left [ W + Y_A +  \tilde{Y}_A  - 2 \, Z \right ]  \,, \nonumber \\
\rho_{n, \rm{LP}}^{(3P)} &=& 2\, (u-1) \, \left ( \Psi_A + \Psi_V \right ) \,, \nonumber \\
\rho_{n, \rm{NLP}}^{(3P)} &=& \left ( \Psi_A - \Psi_V \right )
- \, \left [ X_A +  \tilde{X}_A  - 2 \, Y_A - 2\, \tilde{Y}_A  \right ]  \,, \nonumber \\
\rho_{T, \rm{LP}}^{(3P)} &=&  (1-2 \,u) \, \left ( \Psi_V + X_A  - 2 \, Y_A \right )
+ \Psi_A -  \tilde{X}_A + 2\, \tilde{Y}_A \,, \nonumber \\
\rho_{T, \rm{NLP}}^{(3P)} &=& \left ( \Psi_A - \Psi_V + X_A +  \tilde{X}_A \right )
+ 2 \, \left [ 2\, W +  Y_A  + \, \tilde{Y}_A  - 4 \, Z \right ]  \,,
\end{eqnarray}
where we have suppressed the arguments of the three-particle $B$-meson LCDAs for brevity.
To obtain such three-particle corrections to the  correlation function (\ref{correlator: definition}),
we have adopted the following decomposition of the light-cone matrix element in HQET \cite{Braun:2017liq}
\begin{eqnarray}
&& \langle 0 | \bar q_{\alpha}(z_1 \, \bar n) \, g_s \, G_{\mu \nu}(z_2 \, \bar n) \,
h_{v \, \beta}(0) | \bar B(v) \rangle \nonumber \\
&& = {\tilde{f}_B(\mu) \, m_B \over 4} \,
\bigg [ (1 + \not v) \, \bigg \{ (v_{\mu} \gamma_{\nu} - v_{\nu} \gamma_{\mu})  \,
\left [\Psi_A(z_1, z_2, \mu) - \Psi_V(z_1, z_2, \mu) \right ]
- i \, \sigma_{\mu \nu} \, \Psi_V(z_1, z_2, \mu) \nonumber  \\
&& \hspace{0.4 cm}
- (\bar n_{\mu} \, v_{\nu} - \bar n_{\nu} \, v_{\mu} ) \, X_A(z_1, z_2, \mu)
+ (\bar n_{\mu} \, \gamma_{\nu} - \bar n_{\nu} \, \gamma_{\mu} ) \,
\left [ W(z_1, z_2, \mu)  + Y_A(z_1, z_2, \mu)   \right ] \nonumber \\
&& \hspace{0.4 cm} + \, i \, \epsilon_{\mu \nu \alpha \beta} \,
\bar n^{\alpha} \, v^{\beta}  \, \gamma_5 \, \tilde{X}_A(z_1, z_2, \mu)
- \, i \, \epsilon_{\mu \nu \alpha \beta} \,
\bar n^{\alpha} \, \gamma^{\beta}  \, \gamma_5 \, \tilde{Y}_A(z_1, z_2, \mu)  \nonumber \\
&&  \hspace{0.4 cm}  - \, ( \bar n_{\mu} \, v_{\nu} -  \bar n_{\nu} \, v_{\mu} ) \,
\not \bar  n \, W(z_1, z_2, \mu)
+ \, ( \bar n_{\mu} \, \gamma_{\nu} - \bar n_{\nu} \, \gamma_{\mu} ) \,
\not \bar  n \, Z(z_1, z_2, \mu)   \bigg \}  \, \gamma_5 \bigg ]_{\beta \, \alpha}  \,,
\label{def: 3-particle B-meson DAs}
\end{eqnarray}
where we have neglected the soft Wilson lines to restore the gauge invariance of the
light-ray operator and our convention corresponds to $\epsilon_{0 1 2 3}=-1$.
As emphasized in \cite{Braun:2017liq}, the higher-twist two-particle $B$-meson LCDAs
due to nonvanishing quark transverse momentum can be expressed in terms of the three-particle
configurations with the exact EOM, and they must be taken into account simultaneously
for consistency. Including the light-cone correction terms up to the ${\cal O}(x^2)$ accuracy,
the two-particle renormalized light-cone matrix element (\ref{def: two-particle B-meson DAs})
can be parameterized as follows \cite{Braun:2017liq}
\begin{eqnarray}
&& \langle  0 | \left (\bar d \, Y_s \right)_{\beta} (x) \,
\left (Y_s^{\dag} \, h_v \right )_{\alpha}(0)| \bar B(v)\rangle \nonumber \\
&& = - \frac{i \tilde f_B(\mu) \, m_B}{4}  \,
\int_0^{\infty} \, d \omega \, e^{- i \, \omega \, v \cdot x} \,
\bigg \{  \frac{1+ \! \not v}{2} \, \,
\bigg [ 2 \, \left ( \phi_{B}^{+}(\omega, \mu) + x^2 \, g_B^{+}(\omega, \mu)  \right ) \nonumber \\
&& \hspace{0.5 cm} - {1 \over v \cdot x}  \,
\left  [ \left ( \phi_{B}^{+}(\omega, \mu) - \phi_{B}^{-}(\omega, \mu)  \right )
+  x^2 \, \left ( g_{B}^{+}(\omega, \mu) - g_{B}^{-}(\omega, \mu)  \right )   \right ]  \,
\! \not x \bigg ] \, \gamma_5 \bigg \}_{\alpha \beta}\,,
\label{def: two-particle B-meson DAs with light-cone correction}
\end{eqnarray}
where the two  new distribution amplitudes $g_{B}^{+}$ and $g_{B}^{-}$  are of twist-four and -five, respectively.
Applying the operator identities between the two-body and three-body light-cone operators leads to the nontrivial
relations of  $B$-meson LCDAs in the momentum space
\begin{eqnarray}
- \omega \, {d \over d \omega} \, \phi_B^{-}(\omega, \mu)
&=& \phi_B^{+}(\omega, \mu) - 2 \, \int_0^{\infty} \, {d \omega_2 \over \omega_2^2} \,
\Phi_3(\omega, \omega_2, \mu) + 2 \,  \int_0^{\omega} \, \, {d \omega_2 \over \omega_2^2} \,
\Phi_3(\omega-\omega_2, \omega_2, \mu)  \nonumber \\
&& + \, 2 \,  \int_0^{\omega} \, \, {d \omega_2 \over \omega_2} \,
{d \over d \omega} \, \Phi_3(\omega-\omega_2, \omega_2, \mu) \,,
\label{the first EOM} \\
-2 \, {d^2 \over d \omega^2} \, g_B^{+}(\omega, \mu) &=&
\left [ {3 \over 2} + (\omega - \bar \Lambda) \, {d \over d \omega}   \right ] \, \phi_B^{+}(\omega, \mu)
- {1 \over 2}  \, \phi_B^{-}(\omega, \mu)
+ \int_0^{\infty} \, {d \omega_2 \over \omega_2 } \, {d \over d \omega} \, \Psi_4(\omega, \omega_2, \mu) \nonumber \\
&& - \int_0^{\infty} \, {d \omega_2 \over \omega_2^2 } \, \Psi_4(\omega, \omega_2, \mu)
+ \int_0^{\omega} \, {d \omega_2 \over \omega_2^2 } \, \Psi_4(\omega-\omega_2, \omega_2, \mu) \,,
\label{the second EOM}  \\
-2 \, {d^2 \over d \omega^2} \, g_B^{-}(\omega, \mu) &=&
\left [ {3 \over 2} + (\omega - \bar \Lambda) \, {d \over d \omega}   \right ] \, \phi_B^{-}(\omega, \mu)
- {1 \over 2}  \, \phi_B^{+}(\omega, \mu)
+ \int_0^{\infty} \, {d \omega_2 \over \omega_2 } \, {d \over d \omega} \, \Psi_5(\omega, \omega_2, \mu) \nonumber \\
&& - \int_0^{\infty} \, {d \omega_2 \over \omega_2^2 } \, \Psi_5(\omega, \omega_2, \mu)
+ \int_0^{\omega} \, {d \omega_2 \over \omega_2^2 } \, \Psi_5(\omega-\omega_2, \omega_2, \mu) \,,
\label{the third EOM}  \\
\phi_B^{-}(\omega, \mu) &=&  \left (2 \, \bar \Lambda - \omega \right ) \, {d \phi_B^{+}(\omega, \mu) \over d \omega}
- 2 \, \int_0^{\infty} \, {d \omega_2 \over \omega_2^2 } \, \Phi_4(\omega, \omega_2, \mu)   \nonumber \\
&& + \,  2 \, \int_0^{\omega} \, {d \omega_2 \over \omega_2} \,
\left ( {d \over  d \, \omega_2} +   {d  \over d \, \omega}  \right ) \, \Phi_4(\omega-\omega_2, \omega_2, \mu)  \nonumber \\
&& + \,  2 \,  \int_0^{\omega} \, {d \omega_2 \over \omega_2} \, {d \over d \omega} \, \, \Psi_4(\omega-\omega_2, \omega_2, \mu)
 - \, 2 \,  \int_0^{\infty} \, {d \omega_2 \over \omega_2} \, {d \over d \omega} \, \, \Psi_4(\omega, \omega_2, \mu) \,,
\label{the fourth EOM}
\end{eqnarray}
which can be obtained from the Fourier transformation of the coordinate-space representations obtained in \cite{Braun:2017liq}.
We have introduced the three-particle $B$-meson LCDAs of definite twist
\begin{eqnarray}
\Phi_3(\omega_1, \omega_2, \mu) &=&  \Psi_A(\omega_1, \omega_2, \mu) - \Psi_V(\omega_1, \omega_2, \mu) \,, \nonumber \\
\Phi_4(\omega_1, \omega_2, \mu) &=&  \Psi_A(\omega_1, \omega_2, \mu) + \Psi_V(\omega_1, \omega_2, \mu) \,, \nonumber \\
\Psi_4(\omega_1, \omega_2, \mu) &=&  \Psi_A(\omega_1, \omega_2, \mu) + X_A(\omega_1, \omega_2, \mu) \,, \nonumber \\
\tilde{\Psi}_4(\omega_1, \omega_2, \mu) &=&  \Psi_V(\omega_1, \omega_2, \mu) - \tilde{X}_A(\omega_1, \omega_2, \mu) \,, \nonumber \\
\Phi_5(\omega_1, \omega_2, \mu) &=&  \Psi_A(\omega_1, \omega_2, \mu) + \Psi_V(\omega_1, \omega_2, \mu)
+ 2 \, \left  [ Y_A -  \tilde{Y}_A +  W \right ] (\omega_1, \omega_2, \mu)\,, \nonumber \\
\Psi_5(\omega_1, \omega_2, \mu) &=&  - \Psi_A(\omega_1, \omega_2, \mu) + X_A(\omega_1, \omega_2, \mu)
- 2 \, Y_A(\omega_1, \omega_2, \mu) \,, \nonumber \\
\tilde{\Psi}_5(\omega_1, \omega_2, \mu) &=&  - \Psi_V(\omega_1, \omega_2, \mu) - \tilde{X}_A(\omega_1, \omega_2, \mu)
+ 2 \, \tilde{Y}_A(\omega_1, \omega_2, \mu) \,, \nonumber \\
\Phi_6(\omega_1, \omega_2, \mu) &=&  \Psi_A(\omega_1, \omega_2, \mu) - \Psi_V(\omega_1, \omega_2, \mu)
+ 2 \, \left  [ Y_A  +  \tilde{Y}_A
+  W - 2 \, Z \right ] (\omega_1, \omega_2, \mu)  \,.
\label{3P B-meson DAs of definite twist}
\end{eqnarray}
We are now ready to derive the two-particle higher-twist corrections to the vacuum-to-$B$-meson
correlation function (\ref{correlator: definition}) at tree level
\begin{eqnarray}
\Pi_{\mu, \, V}^{\rm 2PHT} &=& - 4 \, {\tilde{f}_B(\mu) \, m_B \over n \cdot p } \, \bar n_{\mu} \,
\bigg \{ - {1 \over 2} \, \int_0^{\infty} \, d \omega_1 \, \int_0^{\infty} \, d \omega_2 \, \int_0^1 d u \,
{\bar u \, \Psi_5(\omega_1, \omega_2, \mu) \over (\bar n \cdot p - \omega_1 - u \, \omega_2)^2}  \nonumber \\
&& + \int_0^{\infty} \, {d \omega \over (\bar n \cdot p -\omega)^2} \, \hat{g}_B^{-}(\omega, \mu)  \bigg \}  \,,  \nonumber \\
\Pi_{\mu, \, T}^{\rm 2PHT} &=&  2 \,  i \, {\tilde{f}_B(\mu) \, m_B^2 \over n \cdot p } \,
\, \left [ n_{\mu} -  {n \cdot q \over m_B} \, \bar n_{\mu} \right ] \,
\bigg \{ - {1 \over 2} \, \int_0^{\infty} \, d \omega_1 \, \int_0^{\infty} \, d \omega_2 \, \int_0^1 d u \,
{\bar u \, \Psi_5(\omega_1, \omega_2, \mu)  \over (\bar n \cdot p - \omega_1 - u \, \omega_2)^2} \,  \nonumber \\
&& + \int_0^{\infty} \, {d \omega \over (\bar n \cdot p -\omega)^2} \, \hat{g}_B^{-}(\omega, \mu)  \bigg \}  \,,
\end{eqnarray}
where we have introduced the convention
\begin{eqnarray}
\hat{g}_B^{-}(\omega, \mu) =  {1 \over 4} \, \int_{\omega}^{\infty}  \, d \rho \,
\bigg \{ (\rho - \omega) \, \left [ \phi_B^{+}(\rho) -  \phi_B^{-}(\rho) \right ]
- 2 \, (\bar \Lambda - \rho)  \, \phi_B^{-}(\rho) \bigg \}  \,.
\label{def: gBminhat}
\end{eqnarray}

Adding up the two-particle and three-particle higher-twist corrections at tree level together and implementing
the standard strategy to construct the sum rules for heavy-to-light form factors gives rise to  the following
expressions
\begin{eqnarray}
&& {f_{P} \, n \cdot p \over 2} \,\, {\rm exp} \left [- {m_{P}^2 \over n \cdot p \,\, \omega_M} \right ] \,\,
\left [  f_{B \to P}^{+, \, \rm HT}(q^2)  + \frac{m_B} {n \cdot p} \, f_{B \to P}^{0, \, \rm HT}(q^2)  \right ] \, \nonumber \\
&& = - {\tilde{f}_B(\mu) \, m_B \over n \cdot p}  \,
\bigg \{ e^{-\omega_s/\omega_M} \, H_{\bar n, \rm LP}^{\rm 2PHT}(\omega_s, \mu)
+  \int_0^{\omega_s} \, d \omega^{\prime}  \, {1 \over \omega_M} \,
e^{-\omega^{\prime}/\omega_M}  \, H_{\bar n, \rm LP}^{\rm 2PHT}(\omega^{\prime}, \mu) \nonumber \\
&& \hspace{0.4 cm} + \int_0^{\omega_s} \, d \omega_1 \, \int_{\omega_s - \omega_1}^{\infty} \, {d \omega_2 \over \omega_2} \,
e^{-\omega_s/\omega_M} \,
\bigg [ H_{\bar n, \rm LP}^{\rm 3PHT} \left ({\omega_s - \omega_1 \over \omega_2}, \omega_1, \omega_2, \mu \right ) \nonumber \\
&& \hspace{0.8 cm} + {m_q \over n \cdot p} \,
H_{\bar n, \rm NLP}^{\rm 3PHT} \left ({\omega_s - \omega_1 \over \omega_2}, \omega_1, \omega_2, \mu \right ) \bigg ] \nonumber \\
&& \hspace{0.4 cm} + \int_0^{\omega_s} \, d \omega^{\prime} \, \int_0^{\omega^{\prime}} \, d \omega_1 \,
\int_{\omega^{\prime}  - \omega_1}^{\infty} \, {d \omega_2 \over \omega_2} \, {1 \over \omega_M} \, e^{-\omega^{\prime}/\omega_M} \,
\bigg [ H_{\bar n, \rm LP}^{\rm 3PHT} \left ({\omega^{\prime} - \omega_1 \over \omega_2}, \omega_1, \omega_2, \mu \right )
\nonumber \\
&& \hspace{0.8 cm} + {m_q \over n \cdot p} \,H_{\bar n, \rm NLP}^{\rm 3PHT}
\left ({\omega^{\prime} - \omega_1 \over \omega_2}, \omega_1, \omega_2, \mu \right ) \bigg ] \bigg \} \,,
\label{higher twist of fplus}
\\
&& {f_{P} \, n \cdot p \over 2} \,\, {\rm exp} \left [- {m_{P}^2 \over n \cdot p \,\, \omega_M} \right ] \,
{m_B \over n \cdot p -m_B} \,
\left [  f_{B \to P}^{+, \, \rm HT}(q^2)  - \frac{m_B} {n \cdot p} \, f_{B \to P}^{0, \, \rm HT}(q^2)  \right ] \, \nonumber \\
&& = - {\tilde{f}_B(\mu) \, m_B \over n \cdot p}  \,
\bigg \{  \int_0^{\omega_s} \, d \omega_1 \, \int_{\omega_s - \omega_1}^{\infty} \, {d \omega_2 \over \omega_2} \,
e^{-\omega_s/\omega_M} \,
\bigg [ H_{n, \rm LP}^{\rm 3PHT} \left ({\omega_s - \omega_1 \over \omega_2}, \omega_1, \omega_2, \mu \right ) \nonumber \\
&& \hspace{0.8 cm} + {m_q \over n \cdot p} \,
H_{n, \rm NLP}^{\rm 3PHT} \left ({\omega_s - \omega_1 \over \omega_2}, \omega_1, \omega_2, \mu \right ) \bigg ] \nonumber \\
&& \hspace{0.4 cm} + \int_0^{\omega_s} \, d \omega^{\prime} \, \int_0^{\omega^{\prime}} \, d \omega_1 \,
\int_{\omega^{\prime}  - \omega_1}^{\infty} \, {d \omega_2 \over \omega_2} \, {1 \over \omega_M} \, e^{-\omega^{\prime}/\omega_M} \,
\bigg [ H_{n, \rm LP}^{\rm 3PHT} \left ({\omega^{\prime} - \omega_1 \over \omega_2}, \omega_1, \omega_2, \mu \right )
\nonumber \\
&& \hspace{0.8 cm} + {m_q \over n \cdot p} \,H_{n, \rm NLP}^{\rm 3PHT}
\left ({\omega^{\prime} - \omega_1 \over \omega_2}, \omega_1, \omega_2, \mu \right ) \bigg ]  \bigg \} \,,
\label{higher twist of fzero}
\\
&& f_{P} \, n \cdot p \, {\rm exp} \left [- {m_{P}^2 \over n \cdot p \,\, \omega_M} \right ] \,\,
f_{B \to P}^{T, \, \rm HT}(q^2)  \, \nonumber \\
&& = - {\tilde{f}_B(\mu) \, (m_B +m_P) \over n \cdot p}  \,
\bigg \{ e^{-\omega_s/\omega_M} \, H_{T, \rm LP}^{\rm 2PHT}(\omega_s, \mu)
+  \int_0^{\omega_s} \, d \omega^{\prime}  \, {1 \over \omega_M} \,
e^{-\omega^{\prime}/\omega_M}  \, H_{T, \rm LP}^{\rm 2PHT}(\omega^{\prime}, \mu) \nonumber \\
&& \hspace{0.4 cm} + \int_0^{\omega_s} \, d \omega_1 \, \int_{\omega_s - \omega_1}^{\infty} \, {d \omega_2 \over \omega_2} \,
e^{-\omega_s/\omega_M} \,
\bigg [ H_{T, \rm LP}^{\rm 3PHT} \left ({\omega_s - \omega_1 \over \omega_2}, \omega_1, \omega_2, \mu \right ) \nonumber \\
&& \hspace{0.8 cm} + {m_q \over n \cdot p} \,
H_{T, \rm NLP}^{\rm 3PHT} \left ({\omega_s - \omega_1 \over \omega_2}, \omega_1, \omega_2, \mu \right ) \bigg ] \nonumber \\
&& \hspace{0.4 cm} + \int_0^{\omega_s} \, d \omega^{\prime} \, \int_0^{\omega^{\prime}} \, d \omega_1 \,
\int_{\omega^{\prime}  - \omega_1}^{\infty} \, {d \omega_2 \over \omega_2} \, {1 \over \omega_M} \, e^{-\omega^{\prime}/\omega_M} \,
\bigg [ H_{T, \rm LP}^{\rm 3PHT} \left ({\omega^{\prime} - \omega_1 \over \omega_2}, \omega_1, \omega_2, \mu \right )
\nonumber \\
&& \hspace{0.8 cm} + {m_q \over n \cdot p} \,H_{T, \rm NLP}^{\rm 3PHT}
\left ({\omega^{\prime} - \omega_1 \over \omega_2}, \omega_1, \omega_2, \mu \right ) \bigg ] \bigg \} \,,
\label{higher twist of fT}
\end{eqnarray}
where the nonvanishing spectral functions  $H_{i, \rm LP}^{\rm 2PHT}$ and  $H_{i, \rm (N)LP}^{\rm 3PHT}$
($i = n, \, \bar n, \, T$) are given by
\begin{eqnarray}
H_{\bar n, \rm LP}^{\rm 2PHT}(\omega, \mu) &=&
 H_{T, \rm LP}^{\rm 2PHT}(\omega, \mu) = 4 \, \hat{g}_B^{-}(\omega, \mu) \,, \nonumber \\
 H_{n, \rm LP}^{\rm 3PHT} (u, \omega_1, \omega_2, \mu)&=& 2 \, (u-1) \, \Phi_4(\omega_1, \omega_2, \mu)  \,, \nonumber \\
H_{n, \rm NLP}^{\rm 3PHT} (u, \omega_1, \omega_2, \mu)&=& \tilde{\Psi}_5(\omega_1, \omega_2, \mu)
- \Psi_5(\omega_1, \omega_2, \mu)  \,, \nonumber \\
H_{\bar n, \rm LP}^{\rm 3PHT} (u, \omega_1, \omega_2, \mu)&=& \tilde{\Psi}_5(\omega_1, \omega_2, \mu)
- \Psi_5(\omega_1, \omega_2, \mu)  \,, \nonumber \\
H_{\bar n, \rm NLP}^{\rm 3PHT} (u, \omega_1, \omega_2, \mu)&=&  2 \,\Phi_6(\omega_1, \omega_2, \mu)  \,, \nonumber \\
H_{T, \rm LP}^{\rm 3PHT} (u, \omega_1, \omega_2, \mu)&=& 2\, (1-u) \,\Phi_4(\omega_1, \omega_2, \mu)
- \Psi_5(\omega_1, \omega_2, \mu)  + \tilde{\Psi}_5(\omega_1, \omega_2, \mu)  \,, \nonumber \\
H_{T, \rm NLP}^{\rm 3PHT} (u, \omega_1, \omega_2, \mu)&=&  \Psi_5(\omega_1, \omega_2, \mu)
- \tilde{\Psi}_5(\omega_1, \omega_2, \mu)  + 2 \, \Phi_6(\omega_1, \omega_2, \mu)    \,.
\end{eqnarray}
It is evident that the two-particle higher-twist corrections preserve the large-recoil symmetry relations
of the $B \to P$ form factors and the three-particle higher-twist contributions violate such relations
already at tree level (see \cite{Wang:2018wfj} for a similar observation in the context of the $B \to \gamma \ell \nu$ decays).
Employing the power counting scheme for the Borel mass $\omega_M$ and the threshold parameter $\omega_s$ \cite{Wang:2015vgv}
\begin{eqnarray}
\omega_s \sim \omega_M  \sim {\cal O}(\Lambda^2/m_b)\,,
\end{eqnarray}
we can identify the scaling behaviours of the higher-twist corrections to $B \to \pi, K$ form factors
\begin{eqnarray}
f_{B \to P}^{+, \, \rm HT}(q^2) \sim f_{B\to  P}^{0, \, \rm HT}(q^2) \sim
f_{B \to P}^{T, \, \rm HT}(q^2) \sim {\cal O}  \left ( {\Lambda \over m_b} \right )^{5/2}  \,
\end{eqnarray}
in the heavy quark limit, which is suppressed by one power of $\Lambda/m_b$ compared with the leading-twist
contribution in (\ref{NLO sum rules of B to P form factors}).

Collecting different pieces together, the final expressions for the LCSR of $B \to \pi, K$ form factors
at large hadronic recoil can be written as
\begin{eqnarray}
f_{B \to P}^{+}(q^2) &=&  f_{B \to P}^{+, \rm 2PNLL}(q^2) + f_{B \to P}^{+, \, \rm 2PHT}(q^2)
+ f_{B \to P}^{+, \, \rm 3PHT}(q^2)  \,, \nonumber \\
f_{B \to P}^{0}(q^2) &=&  f_{B \to P}^{0, \rm 2PNLL}(q^2) + f_{B \to P}^{0, \, \rm 2PHT}(q^2)
+ f_{B \to P}^{0, \, \rm 3PHT}(q^2)  \,, \nonumber  \\
f_{B \to P}^{T}(q^2) &=&  f_{B \to P}^{T, \rm 2PNLL}(q^2) + f_{B \to P}^{T, \, \rm 2PHT}(q^2)
+ f_{B \to P}^{T, \, \rm 3PHT}(q^2)  \,,
\label{final sum rules}
\end{eqnarray}
where the manifest expressions of $f_{B \to P}^{i, \rm 2PNLL}(q^2)$ ($i=+, \, 0, \, T$) including the light-quark mass effect
can be found in  (\ref{NLO sum rules of B to P form factors}), and
the higher-twist corrections  $f_{B \to P}^{i, \, \rm 2PHT}(q^2)$ and $f_{B \to P}^{i, \, \rm 3PHT}(q^2)$ can be extracted from
(\ref{higher twist of fplus}), (\ref{higher twist of fzero}) and (\ref{higher twist of fT}).

\section{QCD sum rules for the higher-twist $B$-meson LCDAs}
\label{sect: QCDSR for higher-twist DAs}

The objective of this section is to construct a realistic model for the twist-five and -six
$B$-meson LCDAs consistent with the corresponding  asymptotic behaviour at small quark and
gluon momenta, employing the method of QCD sum rules \cite{Grozin:1996pq,Braun:2003wx}.
We introduce the correlation function with two HQET currents
\begin{eqnarray}
F(\omega, z_1, z_2) &=& i \, \int d^4 y \, e^{- i \, \omega \, y} \,
\langle 0 | T  \{ \bar q(z_1 \bar n) \,  Y_s(z_1 \bar n, z_2 \bar n) \,
g_s \, G_{\alpha \beta}(z_2 \bar n) \, Y_s(z_2 \bar n, 0) \, \Gamma_1 \,  h_v(0)\,, \nonumber \\
&& \bar h_v(y \bar n) \,  g_s \, G_{\rho \lambda}(y \bar n) \, \Gamma_2 \, q(y \bar n)  \} |  0 \rangle \,,
\label{correlator for B-meson DAs}
\end{eqnarray}
where the Dirac matrices $\Gamma_1$ and $\Gamma_2$ of the interpolating currents are specified
in Table \ref{table: choices of the interpolating currents}.

\begin{table}[t!bph]
\begin{center}
\begin{tabular}{|c|c|c|}
  \hline
  \hline
 LCDAs & $\Gamma_1$ & $\Gamma_2$ \\
   \hline
  $\Phi_5(\omega_1, \omega_2, \mu)$ & $n^{\beta} \, \not \! \bar n \, \gamma_{\perp}^{\alpha} \, \gamma_5$ & $\sigma^{\rho \lambda} \, \gamma_5$ \\
  $\Psi_5(\omega_1, \omega_2, \mu)$ & $n^{\alpha} \, \bar n^{\beta} \, \not \!  n \,  \gamma_5$ & $\sigma^{\rho \lambda} \, \gamma_5$  \\
  $\tilde{\Psi}_5(\omega_1, \omega_2, \mu)$ & $-{i \over 2} \, \epsilon^{\mu \nu \alpha \beta} \, n_{\mu} \, \bar n_{\nu} \, \not \!  n $ &
  $\sigma^{\rho \lambda} \, \gamma_5$ \\
  $\Phi_6(\omega_1, \omega_2, \mu)$ & $n^{\beta} \, \not \!  n \, \gamma_{\perp}^{\alpha}  \, \gamma_5$ & $n^{\lambda} \, \not \!  n \, \gamma_{\perp}^{\rho}  \, \gamma_5$ \\
  \hline
  \hline
\end{tabular}
\end{center}
\caption{The Dirac structures of the interpolating currents entering the
correlation function (\ref{correlator for B-meson DAs}) and the corresponding
three-particle $B$-meson LCDAs.}
\label{table: choices of the interpolating currents}
\end{table}

Employing the HQET parametrization for the local matrix element with the EOM constraints
for both the heavy and light quarks \cite{Grozin:1996pq}
\begin{eqnarray}
\langle 0 | \bar q \, g_s \, G_{\mu \nu} \, \Gamma \, h_v | \bar B(v) \rangle
&=& - {\tilde{f}_B(\mu) \, m_B \over 6} \,
\bigg \{ i \, \lambda_H^2 {\rm Tr} \left [ \gamma_5 \, \Gamma \, {1 + \not \! v \over 2}  \, \sigma_{\mu \nu} \right ] \nonumber \\
&& + \,  (\lambda_H^2 - \lambda_E^2)  \,{\rm Tr}  \left [ \gamma_5 \, \Gamma \, {1 + \not \! v \over 2}  \,
(v_{\mu} \, \gamma_{\nu} - v_{\nu} \, \gamma_{\mu} ) \right ]  \bigg  \}  \,,
\label{HQET parametrization of the 3P matrix element}
\end{eqnarray}
and comparing (\ref{HQET parametrization of the 3P matrix element}) with the definitions of the three-particle
$B$-meson LCDAs  (\ref{def: 3-particle B-meson DAs}) with the aid of (\ref{3P B-meson DAs of definite twist})
leads to the normalization conditions
\begin{eqnarray}
\Phi_5(z_1=z_2=0, \mu) &=& \int_0^{\infty} \, d \omega_1 \, \int_0^{\infty} \, d \omega_2 \,\, \Phi_5(\omega_1, \omega_2, \mu)
= {\lambda_E^2 + \lambda_H^2 \over 3} \,,  \nonumber \\
\Psi_5(z_1=z_2=0, \mu) &=& \int_0^{\infty} \, d \omega_1 \, \int_0^{\infty} \, d \omega_2 \,\, \Psi_5(\omega_1, \omega_2, \mu)
= - {\lambda_E^2 \over 3} \,,  \nonumber \\
\tilde{\Psi}_5(z_1=z_2=0, \mu) &=& \int_0^{\infty} \, d \omega_1 \, \int_0^{\infty} \, d \omega_2 \,\, \tilde{\Psi}_5(\omega_1, \omega_2, \mu)
= - {\lambda_H^2 \over 3} \,,  \nonumber \\
\Phi_6(z_1=z_2=0, \mu) &=& \int_0^{\infty} \, d \omega_1 \, \int_0^{\infty} \, d \omega_2 \,\, \Phi_6(\omega_1, \omega_2, \mu)
=  {\lambda_E^2 - \lambda_H^2 \over 3} \,.
\label{normalization conditions of twist 5 and 6 DAs}
\end{eqnarray}
The hadronic representation of the HQET correlation function (\ref{correlator for B-meson DAs}) can be written as
\begin{eqnarray}
F(\omega, z_1, z_2)  &=& {1 \over 2 \, (\bar \Lambda - \omega) \, m_B} \,
\langle 0 | \bar q(z_1 \bar n) \,  Y_s(z_1 \bar n, z_2 \bar n) \,
g_s \, G_{\alpha \beta}(z_2 \bar n) \, Y_s(z_2 \bar n, 0) \, \Gamma_1 \,  h_v(0) | \bar B(v) \rangle \nonumber \\
&&  \times \, \langle \bar B(v)  | \bar h_v(0) \,  g_s \, G_{\rho \lambda}(0) \, \Gamma_2 \, q(0)   | 0 \rangle
+ ... \,,
\end{eqnarray}
where $\bar \Lambda=m_B -m_b$ is the effective mass of the $B$-meson in HQET \cite{Falk:1992fm} and the ellipses indicate the
contributions from the higher resonances and continuum states.

\begin{figure}
\begin{center}
\includegraphics[width=0.25 \columnwidth]{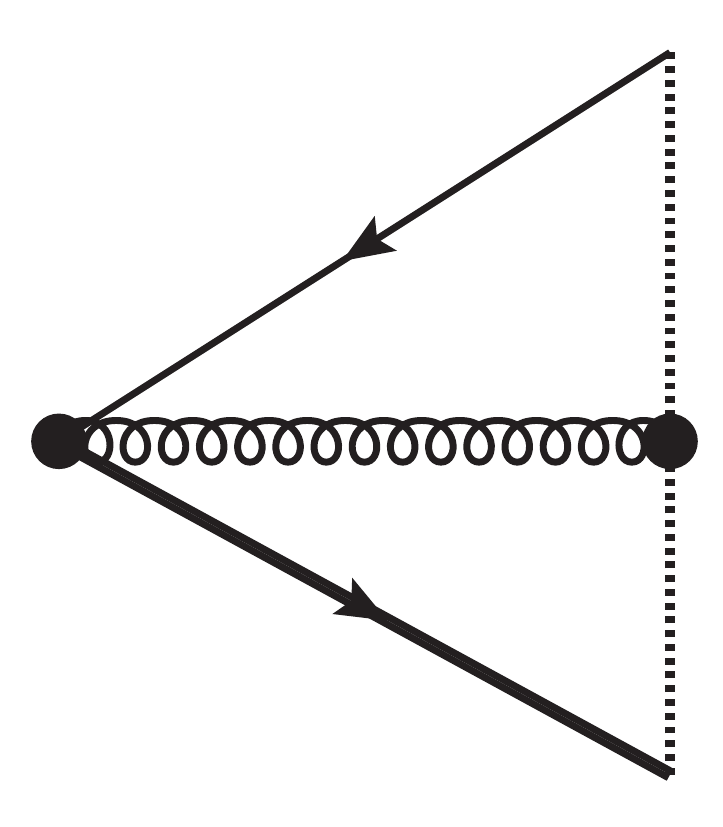}
\vspace*{0.1cm}
\caption{The leading-power contribution to the correlation function
 (\ref{correlator for B-meson DAs}) with two HQET currents at LO in $\alpha_s$.  }
\label{fig: 3P-LCDA-correlator}
\end{center}
\end{figure}

Evaluating the perturbative diagram displayed in figure \ref{fig: 3P-LCDA-correlator} and applying the HQET Feynman rules,
we can readily derive  the leading-power contribution to the correlation function (\ref{correlator for B-meson DAs})
at tree level
\begin{eqnarray}
F(\omega, z_1, z_2) &=& g_s^2 \, C_F \, N_c  \, \int {d^4 k_1 \over (2 \pi)^4} \,
\int {d^4 k_3 \over (2 \pi)^4} \, e^{i \, \bar n \cdot k_1 \, z_1}
\, e^{i \, \bar n \cdot k_3 \, z_2}  \,\,
{\rm Tr} \left [ \not \! k_1  \, \Gamma_1  \, {1 + \not \! v  \over 2} \, \Gamma_2  \right ]  \nonumber \\
&& {1 \over [k_1^2 + i 0] [v \cdot (k_1 +k_3) + \omega + i 0] [k_3^2 + i 0] } \nonumber \\
&& \left [k_{3 \, \alpha}  \, k_{3 \, \rho}  \, g_{\beta \lambda}
- k_{3 \, \alpha}  \, k_{3 \, \lambda}  \, g_{\beta \rho}
- k_{3 \, \beta}  \, k_{3 \, \rho}  \, g_{\alpha \lambda}
+ k_{3 \, \beta}  \, k_{3 \, \lambda}  \, g_{\alpha \rho}  \right ] \,.
\label{QCD result of the correlator for 3P DAs}
\end{eqnarray}
Performing the loop momentum integration in (\ref{QCD result of the correlator for 3P DAs})
and matching the two different representations of the correlation function (\ref{correlator for B-meson DAs})
with the parton-hadron duality ansatz,  we obtain the QCD sum rules for the three-particle higher-twist
$B$-meson LCDAs
\begin{eqnarray}
&& [\tilde{f}_B(\mu)]^2  \, m_B \, (\lambda_H^2 + \lambda_E^2) \,
\Phi_5(\omega_1, \omega_2, \mu) \,\nonumber \\
&&  = - {g_s^2 \, C_F \, N_c \over 96 \, \pi^4} \,
\int_{\omega_1 + \omega_2 \over 2}^{\omega_0} \, ds \,
{\rm exp} \left [{\bar \Lambda - s \over \omega_M}  \right ]  \,
\omega_1 \, (\omega_1 + \omega_2 - 2 \, s)^3 \, \theta(2 \, s - \omega_1 - \omega_2)   \,,
\nonumber \\
&& [\tilde{f}_B(\mu)]^2  \, m_B \, (\lambda_H^2 + \lambda_E^2) \,
\Psi_5(\omega_1, \omega_2, \mu) \,\nonumber \\
&&  =  {g_s^2 \, C_F \, N_c \over 192 \, \pi^4} \,
\int_{\omega_1 + \omega_2 \over 2}^{\omega_0} \, ds \,
{\rm exp} \left [{\bar \Lambda - s \over \omega_M}  \right ]  \,
\omega_2 \, (\omega_1 + \omega_2 - 2 \, s)^3 \, \theta(2 \, s - \omega_1 - \omega_2)   \,,
\nonumber \\
&& [\tilde{f}_B(\mu)]^2  \, m_B \, (\lambda_H^2 + \lambda_E^2) \,
\tilde{\Psi}_5(\omega_1, \omega_2, \mu) \,\nonumber \\
&&  =  {g_s^2 \, C_F \, N_c \over 192 \, \pi^4} \,
\int_{\omega_1 + \omega_2 \over 2}^{\omega_0} \, ds \,
{\rm exp} \left [{\bar \Lambda - s \over \omega_M}  \right ]  \,
\omega_2 \, (\omega_1 + \omega_2 - 2 \, s)^3 \, \theta(2 \, s - \omega_1 - \omega_2)   \,,
\nonumber \\
&& [\tilde{f}_B(\mu)]^2  \, m_B \, (\lambda_E^2 - \lambda_H^2) \,
\Phi_6(\omega_1, \omega_2, \mu) \,\nonumber \\
&&  =  {g_s^2 \, C_F \, N_c \over 128 \, \pi^4} \,
\int_{\omega_1 + \omega_2 \over 2}^{\omega_0} \, ds \,
{\rm exp} \left [{\bar \Lambda - s \over \omega_M}  \right ]  \,
(\omega_1 + \omega_2 - 2 \, s)^4 \, \theta(2 \, s - \omega_1 - \omega_2)   \,,
\label{QCDSR for 3P B-meson DAs}
\end{eqnarray}
where the Borel transformation with respect to the variable $\omega$ have been implemented to suppress the
higher-order nonperturbative corrections and minimize the model dependence on the continuum contributions.
For the phenomenological applications, we first suggest the local duality model for the three-particle
$B$-meson LCDAs by taking  the limit $\omega_M \to \infty$ of the obtained QCD sum rules (\ref{QCDSR for 3P B-meson DAs})
\begin{eqnarray}
\Phi_5^{\rm LD}(\omega_1, \omega_2, \mu)  &=& {35 \over 64} \, (\lambda_E^2 + \lambda_H^2) \,
{\omega_1 \over \omega_0^7} \, (2 \, \omega_0 - \omega_1-\omega_2)^4 \,
\theta(2 \, \omega_0 - \omega_1 - \omega_2)  \,, \nonumber \\
\Psi_5^{\rm LD}(\omega_1, \omega_2, \mu)  &=& - {35 \over 64} \, \lambda_E^2 \,
{\omega_2 \over \omega_0^7} \, (2 \, \omega_0 - \omega_1-\omega_2)^4 \,
\theta(2 \, \omega_0 - \omega_1 - \omega_2)  \,, \nonumber \\
\tilde{\Psi}_5^{\rm LD}(\omega_1, \omega_2, \mu)  &=& - {35 \over 64} \, \lambda_H^2 \,
{\omega_2 \over \omega_0^7} \, (2 \, \omega_0 - \omega_1-\omega_2)^4 \,
\theta(2 \, \omega_0 - \omega_1 - \omega_2)  \,, \nonumber \\
\Phi_6^{\rm LD}(\omega_1, \omega_2, \mu)  &=& {7 \over 64} \, (\lambda_E^2 - \lambda_H^2) \,
{1 \over \omega_0^7} \, (2 \, \omega_0 - \omega_1-\omega_2)^5 \,
\theta(2 \, \omega_0 - \omega_1 - \omega_2)  \,.
\label{LD model for the 3P B-meson DAs}
\end{eqnarray}
It is then straightforward to verify that the asymptotic behaviour of the twist-five and -six
$B$-meson LCDAs from the local duality model (\ref{LD model for the 3P B-meson DAs})
\begin{eqnarray}
\Phi_5(\omega_1, \omega_2, \mu) \sim \omega_1 \,, \qquad
\Psi_5(\omega_1, \omega_2, \mu)  \sim
\tilde{\Psi}_5(\omega_1, \omega_2, \mu) \sim \omega_2 \,,
\qquad
\Phi_6(\omega_1, \omega_2, \mu) \sim 1 \,,
\label{asymtotic behaviour of twist 5 and 6 DAs}
\end{eqnarray}
in agreement with the predictions from the RG equations at one loop \cite{Braun:1989iv}.
We further present the local duality model for the remaining two-particle and three-particle $B$-meson LCDAs
constructed in \cite{Braun:2017liq}
\begin{eqnarray}
\phi_B^{+, \rm LD}(\omega, \mu) &=& {5 \over 8 \, \omega_0^5}  \, \omega(2 \, \omega_0 - \omega)^3 \,
\theta(2 \, \omega_0 - \omega)\,,  \nonumber  \\
\phi_B^{-, \rm LD}(\omega, \mu) &=& {5 (2 \, \omega_0 - \omega)^2 \over 192\, \omega_0^5}  \,
\bigg \{6 \, (2 \, \omega_0 - \omega)^2 - {7 \, (\lambda_E^2 - \lambda_H^2) \over \omega_0^2} \,
(15 \, \omega^2 - 20 \, \omega \, \omega_0 + 4 \, \omega_0^2) \bigg \}   \, \nonumber \\
&& \times \, \theta(2 \, \omega_0 - \omega)\,,  \nonumber  \\
\Phi_3^{\rm LD}(\omega_1, \omega_2, \mu) &=&  {105 \, (\lambda_E^2 - \lambda_H^2) \over 8 \, \omega_0^7} \,
\omega_1 \, \omega_2^2 \, \left (\omega_0 - {\omega_1 + \omega_2 \over 2} \right )^2  \,
\theta(2 \, \omega_0 - \omega_1 - \omega_2)  \,,  \nonumber  \\
\Phi_4^{\rm LD}(\omega_1, \omega_2, \mu) &=&  {35 \, (\lambda_E^2 + \lambda_H^2) \over 4 \, \omega_0^7} \,
\omega_2^2 \, \left (\omega_0 - {\omega_1 + \omega_2 \over 2} \right )^3  \,
\theta(2 \, \omega_0 - \omega_1 - \omega_2)  \,,  \nonumber  \\
\Psi_4^{\rm LD}(\omega_1, \omega_2, \mu) &=&  {35 \, \lambda_E^2 \over 2 \, \omega_0^7} \,
\omega_1 \, \omega_2 \, \left (\omega_0 - {\omega_1 + \omega_2 \over 2} \right )^3  \,
\theta(2 \, \omega_0 - \omega_1 - \omega_2)  \,,  \nonumber  \\
\tilde{\Psi}_4^{\rm LD}(\omega_1, \omega_2, \mu) &=&  {35 \, \lambda_H^2 \over 2 \, \omega_0^7} \,
\omega_1 \, \omega_2 \, \left (\omega_0 - {\omega_1 + \omega_2 \over 2} \right )^3  \,
\theta(2 \, \omega_0 - \omega_1 - \omega_2)  \,,
\end{eqnarray}
from which we can further derive the corresponding model for the ``effective" distribution amplitude
defined in (\ref{def: gBminhat})
\begin{eqnarray}
\hat{g}_B^{-, \rm LD}(\omega, \mu) &=& {\omega \, (2 \, \omega_0 - \omega)^3 \over \omega_0^5} \,
\left \{ {5 \over 256} \, (2 \, \omega_0 - \omega)^2  - {35 \, (\lambda_E^2 - \lambda_H^2) \over 1536} \,
\left [ 4 - 12 \, \left ( {\omega \over \omega_0} \right )
+ 11 \, \left ( {\omega \over \omega_0} \right )^2  \right ]\right \}  \nonumber \\
&& \times \, \theta(2 \, \omega_0 - \omega) \,.
\end{eqnarray}
Applying the EOM constraint between the leading-twist and the higher-twist $B$-meson LCDAs (\ref{the fourth EOM}),
the HQET parameters entering  the local duality model for the $B$-meson LCDAs must satisfy the following
relations \cite{Braun:2017liq}
\begin{eqnarray}
\omega_0 = {5 \over 2} \, \lambda_B = 2 \, \bar \Lambda\,, \qquad
3 \, \omega_0^2 = 14 \, (2 \, \lambda_E^2 + \lambda_H^2)\,.
\label{HQET relations for LD model}
\end{eqnarray}

An alternative model for the twist-five and -six $B$-meson LCDAs
consistent with the asymptotic behaviours (\ref{asymtotic behaviour of twist 5 and 6 DAs})
and the normalization conditions (\ref{normalization conditions of twist 5 and 6 DAs})
can be constructed by implementing an exponential falloff at large quark and gluon momenta
\begin{eqnarray}
\Phi_5^{\rm exp}(\omega_1, \omega_2, \mu)
&=& {\lambda_E^2 + \lambda_H^2 \over 3 \, \omega_0^3} \, \omega_1 \,
e^{-(\omega_1 + \omega_2)/\omega_0} \,, \nonumber \\
\Psi_5^{\rm exp}(\omega_1, \omega_2, \mu)
&=& - {\lambda_E^2 \over 3 \, \omega_0^3} \, \omega_2 \,
e^{-(\omega_1 + \omega_2)/\omega_0} \,, \nonumber \\
\tilde{\Psi}_5^{\rm exp}(\omega_1, \omega_2, \mu)
&=& - {\lambda_H^2 \over 3 \, \omega_0^3} \, \omega_2 \,
e^{-(\omega_1 + \omega_2)/\omega_0} \,, \nonumber \\
\Phi_6^{\rm exp}(\omega_1, \omega_2, \mu)
&=& {\lambda_E^2 - \lambda_H^2 \over 3 \, \omega_0^2} \,
e^{-(\omega_1 + \omega_2)/\omega_0} \,.
\end{eqnarray}
We further collect the exponential model for the remaining LCDAs obtained in \cite{Braun:2017liq}
\begin{eqnarray}
\phi_B^{+, \, \rm exp}(\omega, \mu) &=& {\omega \over \omega_0^2} \,  e^{-\omega/\omega_0} \,, \nonumber \\
\phi_B^{-, \, \rm exp}(\omega, \mu) &=& {1 \over \omega_0} \,  e^{-\omega/\omega_0}
- {\lambda_E^2 - \lambda_H^2 \over 9 \, \omega_0^3}  \,
\left [ 1 - 2 \, \left ( {\omega \over \omega_0} \right )
+ {1 \over 2} \, \left ( {\omega \over \omega_0} \right )^2 \right ]
\,  e^{-\omega/\omega_0} \,, \nonumber \\
\Phi_3^{\rm exp}(\omega_1, \omega_2, \mu) &=&
{\lambda_E^2 - \lambda_H^2 \over 6 \, \omega_0^5} \, \omega_1 \, \omega_2^2 \,
e^{-(\omega_1 + \omega_2)/\omega_0} \,, \nonumber \\
\Phi_4^{\rm exp}(\omega_1, \omega_2, \mu) &=&
{\lambda_E^2 + \lambda_H^2 \over 6 \, \omega_0^4} \, \omega_2^2 \,
e^{-(\omega_1 + \omega_2)/\omega_0} \,, \nonumber \\
\Psi_4^{\rm exp}(\omega_1, \omega_2, \mu) &=&
{\lambda_E^2 \over 3 \, \omega_0^4} \, \omega_1 \, \omega_2 \,
e^{-(\omega_1 + \omega_2)/\omega_0} \,, \nonumber \\
\tilde{\Psi}_4^{\rm exp}(\omega_1, \omega_2, \mu) &=&
{\lambda_H^2 \over 3 \, \omega_0^4} \, \omega_1 \, \omega_2 \,
e^{-(\omega_1 + \omega_2)/\omega_0} \,,
\end{eqnarray}
which imply the following expression for the two-particle twist-five  LCDA
\begin{eqnarray}
\hat{g}_B^{-,  \, \rm exp}(\omega, \mu) = \omega \,
\left \{ {3 \over 4}  - {\lambda_E^2 - \lambda_H^2 \over 12 \, \omega_0^2} \,
\left [ 1 - \left ( {\omega \over \omega_0} \right )
+ {1 \over 3} \, \left ( {\omega \over \omega_0} \right )^2 \right ] \right \}
\,  e^{-\omega/\omega_0}  \,.
\end{eqnarray}
Implementing the EOM constraint (\ref{the fourth EOM}) for the exponential model leads to
\begin{eqnarray}
\omega_0 =  \lambda_B = {2 \over 3} \, \bar \Lambda\,, \qquad
2 \, \bar \Lambda^2 = 2 \, \lambda_E^2 + \lambda_H^2 \,.
\label{HQET relations for exponential model}
\end{eqnarray}
It is worthwhile to point that the HQET relations (\ref{HQET relations for LD model}) and
(\ref{HQET relations for exponential model}) are derived from the classical EOM with the
assumption that the first two moments of the leading-twist $B$-meson LCDA $\phi_B^{+}(\omega, \mu)$
are finite. Apparently, perturbative QCD corrections to the $B$-meson LCDAs will
violate such tree-level relations in a nontrivial way (see \cite{Feldmann:2014ika} for further discussion).

\section{Numerical analysis}
\label{sect: numerical analysis}

The purpose of this section is to explore  phenomenological implications of the newly derived sum rules
(\ref{final sum rules}) for $B \to \pi, K$ form factors with the subleading-twist corrections.
We will place particular attention to the normalized differential $q^2$ distributions of $B \to \pi \ell \nu_{\ell}$
($\ell=\mu\,, \tau$), the determination of the CKM matrix element $|V_{ub}|$ and the differential branching fractions
of the rare exclusive $B \to K \nu \nu$ decays.

\subsection{Theory inputs}
\label{section: theory inputs}

We will proceed by specifying  the theory inputs entering the LCSR for $B \to \pi, K$ form factors,
including the shape parameters of $B$-meson LCDAs, the intrinsic sum rule parameters and
the decay constants  of the $B$-meson and the light pseudoscalar mesons.
Due to the EOM constraints (\ref{HQET relations for LD model}) and
(\ref{HQET relations for exponential model}),  only two of the three HQET parameters $\lambda_B(\mu)$,
$\lambda_E(\mu)$ and $\lambda_H(\mu)$ appearing in the $B$-meson LCDAs  are independent of each other.
As observed in \cite{Braun:2017liq}, the ratio $R(\mu)=\lambda^2_E(\mu)/\lambda^2_H(\mu)$ estimated from the QCD sum rule
approach \cite{Grozin:1996pq,Nishikawa:2011qk} is insensitive to the perturbative QCD corrections and
the higher-order nonperturbative QCD corrections.
We will therefore take $\lambda_B(\mu)$ and $R(\mu)$ as free parameters in the numerical analysis.
The renormalization scale dependence of the inverse moment $\lambda_B(\mu)$
\begin{eqnarray}
\lambda_B(\mu)= \lambda_B(\mu_0) \, \left \{ 1 +  {\alpha_s(\mu_0) \, C_F \over 4 \, \pi} \,
\ln {\mu \over \mu_0}  \, \left [  2 - 2 \, \ln {\mu \over \mu_0}  - 4 \, \sigma_1(\mu_0) \right ]
+ {\cal O}(\alpha_s^2)  \right \}^{-1}    \,
\end{eqnarray}
can be obtained from the Lange-Neubert evolution equation of $\phi_B^{+}(\omega, \mu)$ \cite{Lange:2003ff}.
We employ the definition of the inverse-logarithmic moment
\begin{eqnarray}
\sigma_1(\mu) = \lambda_B(\mu) \, \int_0^{\infty} \, {d \omega \over \omega}  \,
\ln {\mu \over \omega}  \, \phi_B^{+}(\omega, \mu) \,
\end{eqnarray}
and adopt the interval $\sigma_1(\mu_0) = 1.4 \pm 0.4$  from the QCD sum rule calculation
for $\mu_0=1 \, {\rm GeV}$ \cite{Braun:2003wx}.
The ratio $R(\mu_0) =0.5 \pm 0.1$ based upon the nonperturbative QCD  computations
\cite{Grozin:1996pq,Nishikawa:2011qk} will be taken in the subsequent calculations.

Implementing the standard procedure for the determinations of the internal sum rule
parameters as discussed in \cite{Wang:2015vgv} gives rise to
\begin{eqnarray}
M^2 = n \cdot p \, \omega_M &=& (1.25 \pm 0.25) \, {\rm GeV^2}\,,
\qquad s_0^{\pi} = n \cdot p  \, \omega_s^{\pi} = (0.70 \pm 0.05) \, {\rm GeV^2} \,, \nonumber \\
s_0^{K} = n \cdot p  \, \omega_s^{K} &=& (1.05 \pm 0.05) \, {\rm GeV^2}  \,,
\end{eqnarray}
in agreement with the values used for the LCSR of the pion-photon form factor \cite{Wang:2017ijn}
and for the two-point QCD sum rules of the kaon decay constant \cite{Khodjamirian:2003xk}.

By virtue of the matching relation (\ref{HQET matching of fB}),  the HQET decay constant $\tilde{f}_B(\mu)$
will be related to the QCD decay constant $f_B$, for which we will take the averaged Lattice results
$f_B=(192.0 \pm 4.3) \, {\rm MeV}$ \cite{Aoki:2016frl}  with $N_f=2+1$.
In addition, the QCD decay constants of the light pseudoscalar mesons
\begin{eqnarray}
f_{\pi}=(130.2 \pm 1.7) \, {\rm MeV} \,, \qquad
f_{K}=(155.6 \pm 0.4) \, {\rm MeV}  \,
\end{eqnarray}
are borrowed from the Particle Data Group (PDG) \cite{Tanabashi:2018oca},
which differ  slightly  from the Flavour Lattice Averaging Group (FLAG) values \cite{Aoki:2016frl}
mainly due to the  different treatments of theory uncertainties and correlations.

The masses of the light quarks in the $\overline{\rm MS}$ scheme summarized in PDG \cite{Tanabashi:2018oca}
\begin{eqnarray}
m_u(2 \, {\rm GeV}) &=& (2.15 \pm 0.15) \, {\rm MeV}  \,, \qquad
m_d(2 \, {\rm GeV}) = (4.70 \pm 0.20) \, {\rm MeV}  \,, \nonumber \\
m_s(2 \, {\rm GeV}) &=& (93.8 \pm 1.5 \pm 1.9) \, {\rm MeV}  \,,
\end{eqnarray}
will be employed in the following. We further take the numerical values of the
$\overline{\rm MS}$ bottom quark mass  determined from non-relativistic sum rules \cite{Beneke:2014pta}
(see \cite{Dehnadi:2015fra} for independent determinations from relativistic sum rules
with similar results)
\begin{eqnarray}
\overline{m_b}(\overline{m_b}) = (4.193^{+0.022}_{-0.035}) \, {\rm GeV} \,.
\end{eqnarray}

Following the discussion presented in \cite{Wang:2015vgv}, the factorization scale
entering the leading-twist LCSR for $B \to \pi, K$ form factors at NLL will be varied
in the interval $1 \, {\rm GeV} \leq \mu \leq 2 \, {\rm GeV}$ around the default value
$\mu=1.5 \, {\rm GeV}$. The hard scales $\mu_{h1}$ and $\mu_{h2}$ as well as the QCD renormalization
scale for the tensor current $\nu_h$ will be taken as
$\mu_{h1}=\mu_{h2}=\nu_h=m_b$ with the variation in the range $[m_b/2, 2 \, m_b]$.

\subsection{Predictions for $B \to \pi, K$ form factors }

We will proceed to investigate the numerical impacts of the higher-twist corrections
and the SU(3)-flavour symmetry breaking effects computed from the method of LCSR.
Prior to presenting the breakdown of the distinct terms contributing to
the semileptonic $B \to \pi, K$ decay form factors, we need to determine the inverse moment
$\lambda_B(\mu_0)$ of the leading-twist $B$-meson LCDA $\phi_B^{+}(\omega, \mu)$.
Despite of the numerous  studies of $\lambda_B(\mu_0)$ with the direct nonperturbative
calculations \cite{Braun:2003wx} and the indirect determinations from measurements
of the partial branching fractions of $B \to \gamma \ell \nu$
\cite{Ball:2003fq,Wang:2018wfj,Braun:2012kp,Wang:2016qii,Beneke:2018wjp},
the current constraints of $\lambda_B(\mu_0)$ are still far from satisfactory
due to the systematic uncertainty of the direct QCD approach and the sensitivity
of the $B \to \gamma$ form factors to the shape of $\phi_B^{+}(\omega, \mu)$ at small $\omega$.
Following the strategy displayed in \cite{Wang:2015vgv}, we will match our calculations for the
vector $B \to \pi$ form factor at $q^2=0$ from the LCSR with $B$-meson LCDAs
with the independent predictions $f_{B \to \pi}^{+}(q^2=0)=0.28 \pm 0.03$ from the LCSR with pion LCDAs including the higher-twist
corrections up to twist-four accuracy \cite{Khodjamirian:2011ub}
(see \cite{Imsong:2014oqa,Khodjamirian:2017fxg} for slightly different values).
Performing such matching procedure we obtain
\begin{eqnarray}
\lambda_B(\mu_0) = \left\{
\begin{array}{l}
285^{+27}_{-23} \,\, {\rm MeV}  \,, \qquad  \hspace{1.5 cm}
(\rm Exponential \,\, Model) \vspace{1.0 cm} \\
286^{+26}_{-22} \,\, {\rm MeV}  \,.
 \qquad  \hspace{1.5 cm}
(\rm Local \,\, Duality \,\, Model)
\end{array}
 \hspace{0.5 cm} \right.
\,
\label{values of lambdaB}
\end{eqnarray}
It is apparent that the determined values of $\lambda_B(\mu_0)$ for the considered two models of
$B$-meson LCDAs are practically identical, which can be understood from the fact that the small $\omega$
behaviours of the above-mentioned two models are very similar to each other
(albeit with the rather different high-energy behaviours) as observed in \cite{Braun:2017liq}.
The determined values of $\lambda_B(\mu_0)$ (\ref{values of lambdaB}) differ from the previous interval
presented in \cite{Wang:2015vgv}, where only the leading-power two-particle contributions to the sum rules
were taken into account at NLL.
For the illustration purpose,  we will adopt the exponential model for  $B$-meson LCDAs as our default choice
and the theory uncertainty due to the model dependence of these distribution amplitudes will be included in
the final predictions for  $B \to \pi, K$ form factors.

\begin{figure}
\begin{center}
\includegraphics[width=0.55 \columnwidth]{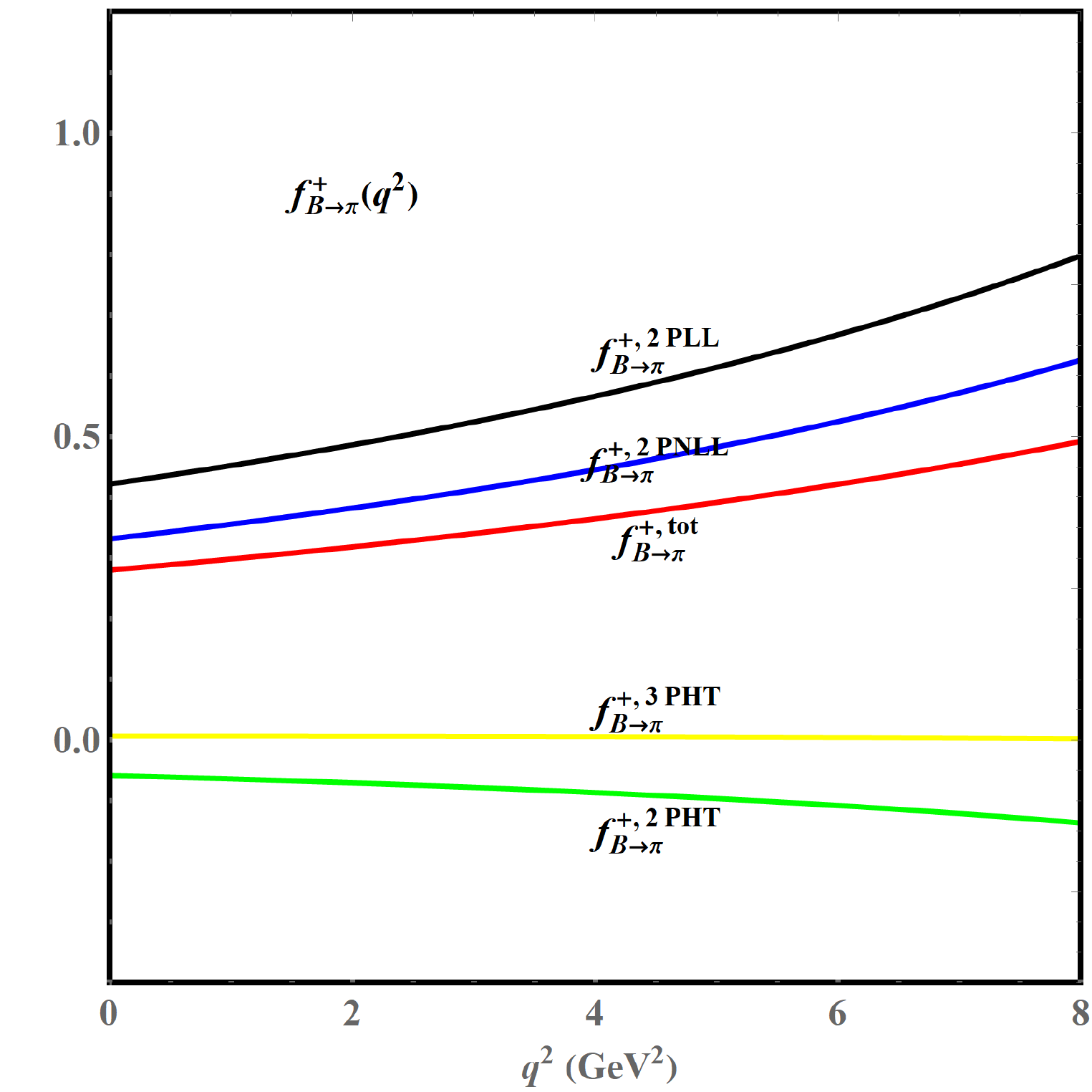}
\vspace*{0.1cm}
\caption{The momentum-transfer dependence of the vector $B \to \pi$ form factor
from the leading-power contribution at LL ($f_{B \to \pi}^{+, \, \rm 2PLL}$, black),
the leading-power contribution  at NLL ($f_{B \to \pi}^{+, \, \rm 2PNLL}$, blue),
the two-particle higher-twist correction  ($f_{B \to \pi}^{+, \, \rm 2PHT}$, green),
and the three-particle higher-twist correction ($f_{B \to \pi}^{+, \, \rm 3PHT}$, yellow).}
\label{fig: Breakdown of the vector B to pi form factor}
\end{center}
\end{figure}

\begin{figure}
\begin{center}
\includegraphics[width=0.45 \columnwidth]{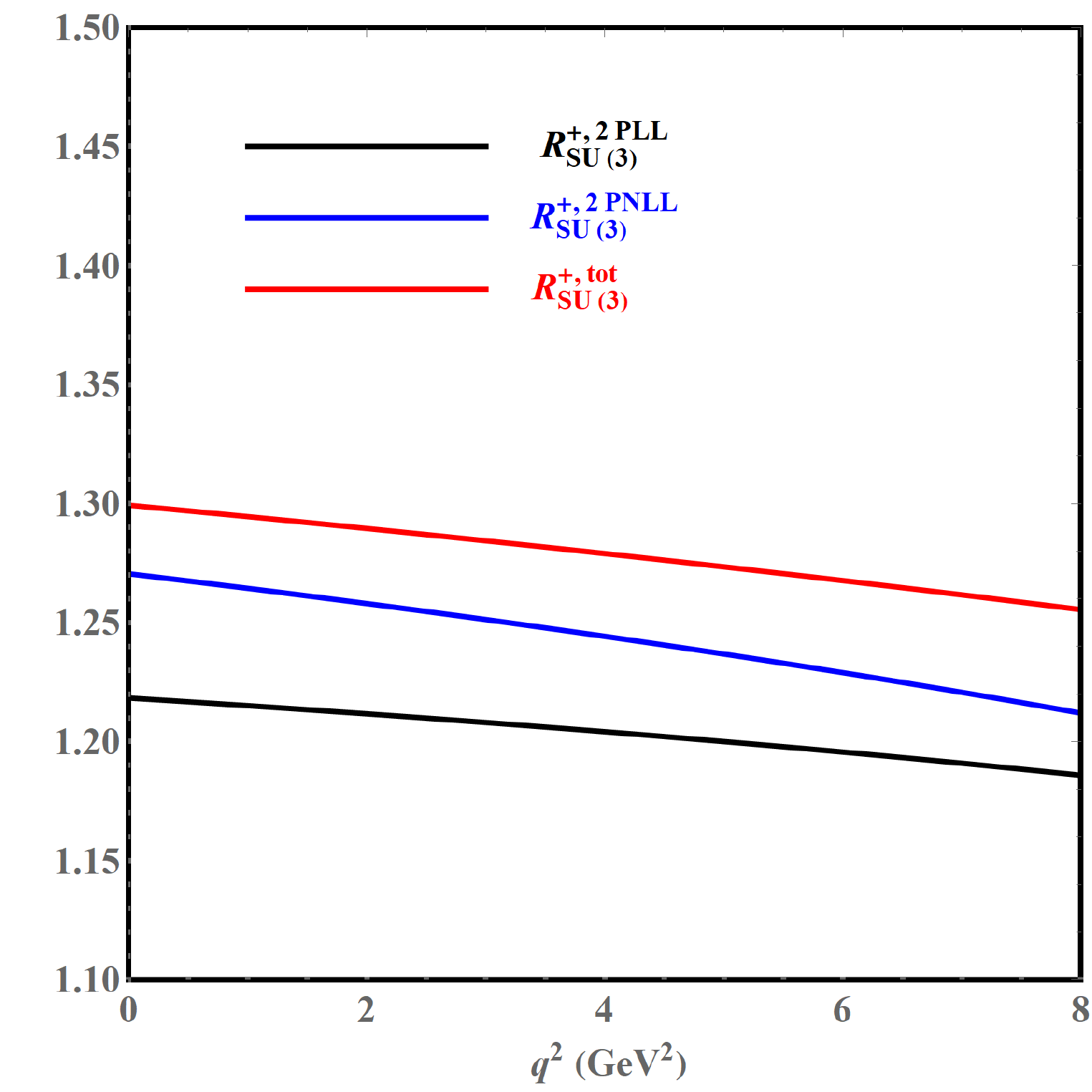}
\hspace{0.5 cm}
\includegraphics[width=0.45 \columnwidth]{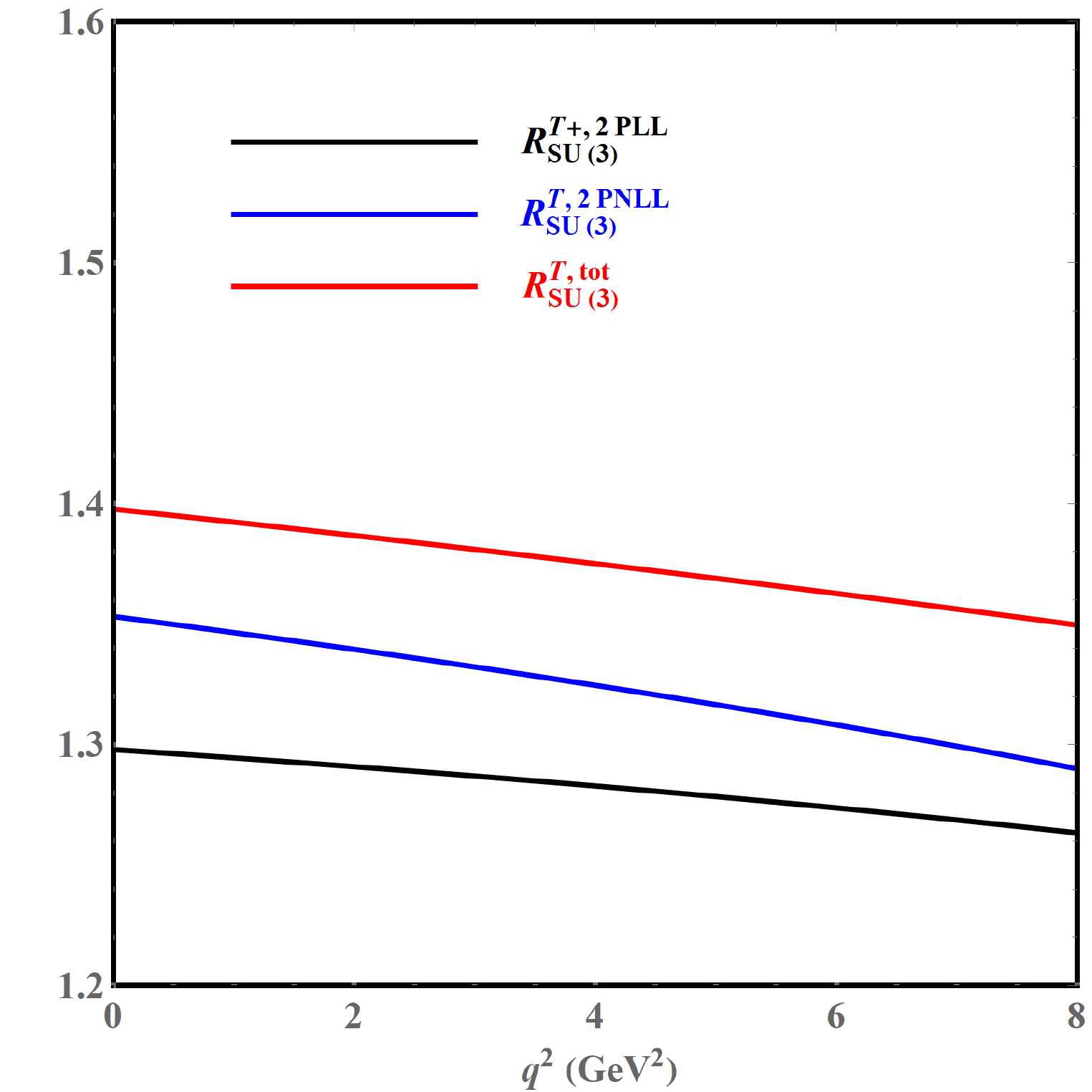}
\vspace*{0.1cm}
\caption{The SU(3)-flavour symmetry breaking effects between
$B \to \pi$ and $B \to K$ form factors predicted from the LCSR with
$B$-meson LCDAs. The momentum-transfer dependence of the ratio
$R_{\rm SU(3)}^{0} (q^2)$ behaves in a similar way to $R_{\rm SU(3)}^{+} (q^2)$
at $0 \leq q^2 \leq 8 \, {\rm GeV^2}$ and we will therefore not display this ratio for brevity. }
\label{fig: SU(3) breaking for B to P form factors}
\end{center}
\end{figure}

We first display the breakdown of distinct pieces contributing to the LCSR of  the vector $B \to \pi$ form factor
at $0 \leq q^2 \leq 8 \, {\rm GeV^2}$ in figure \ref{fig: Breakdown of the vector B to pi form factor}.
It is evident that higher-twist corrections to the $B \to \pi$ form factor
$f_{B \to \pi}^{+}(q^2)$ are dominated by the two-particle twist-five contribution from $\hat{g}_B^{-}(\omega, \mu)$,
which can shift the leading-power prediction by an amount of approximately $(20 \sim 30) \%$.
The three-particle higher-twist contribution only generates a minor impact on the theory prediction of
$f_{B \to \pi}^{+}(q^2)$ and numerically ${\cal O}(2 \%)$.
We further observe that the NLL QCD correction to the leading-power contribution can yield
approximately ${\cal O}(20 \%)$ reduction of the corresponding LL QCD prediction.
We have also verified that such observations also hold true for the momentum-transfer dependence of
the scalar and tensor $B \to \pi$ form factors at large hadronic recoil.
The SU(3)-flavour symmetry breaking effects between the $B \to \pi$ and $B \to K$ form factors
\begin{eqnarray}
R_{\rm SU(3)}^{i} (q^2) =  \frac{f_{B \to K}^{i}(q^2)}{f_{B \to \pi}^{i}(q^2)} \,, \qquad
({\rm with}  \,\, i=+, \, 0, \, T) \,
\end{eqnarray}
which originate from the nonvanishing strange-quark mass,
from the discrepancy between the threshold parameters for the pion and kaon channels
and from the difference between the decay constants $f_{\pi}$ and $f_K$ are presented
in figure \ref{fig: SU(3) breaking for B to P form factors}.
It can be observed that our predictions for the
SU(3)-flavour symmetry breaking effects are in good agreement with that obtained from the LCSR with
the light-meson LCDAs \cite{Khodjamirian:2017fxg}, but are somewhat smaller
than the previous calculations \cite{Duplancic:2008tk}.

We are now in a position to discuss the large-recoil symmetry breaking effects of $B \to P$
form factors due to both the perturbative QCD corrections at leading power in $1/m_b$
and the subleading power soft contributions from the three-particle higher-twist $B$-meson LCDAs.
To compare our predictions with the perturbative calculations from the QCD factorization approach,
we collect the factorization formulae for the heavy-to-light $B$-meson form factors
in the heavy quark limit at one loop \cite{Beneke:2000wa} (see \cite{Beneke:2005gs,Bell:2010mg} for further improvement)
\begin{eqnarray}
f_{B \to P}^{0}(q^2) &=& {n \cdot p \over m_B} \, f_{B \to P}^{+}(q^2)  \,
\left [  1 +  {\alpha_s \, C_F \over 2 \, \pi} \,
\left (1 -  { n \cdot p \over n \cdot p - m_B} \, \ln {n \cdot p \over m_B} \right ) \right ] \, \nonumber \\
&& +  \,  {m_B - n \cdot p \over n \cdot p } \, {\alpha_s \, C_F \over 4 \, \pi} \,
{8\, \pi^2 \, f_B \, f_P \over N_c \, m_B}  \, \int_0^1 \, d u \, { \phi_P(u, \mu) \over \bar u} \,
\int_0^{\infty} \, d \omega \, {\phi_B^{+}(\omega, \mu) \over \omega} \,,
\\
f_{B \to P}^{T}(q^2) &=& {m_B + m_P \over m_B} \, f_{B \to P}^{+}(q^2)  \,
\left [  1 +  {\alpha_s \, C_F \over 4 \, \pi} \,
\left ( \ln {m_b^2 \over \mu^2} + 2\,  { n \cdot p \over n \cdot p - m_B} \,
\ln {n \cdot p \over m_B} \right ) \right ] \, \nonumber \\
&& -  \,  {m_B + m_P \over n \cdot p } \, {\alpha_s \, C_F \over 4 \, \pi} \,
{8\, \pi^2 \, f_B \, f_P \over N_c \, m_B}  \, \int_0^1 \, d u \, { \phi_P(u, \mu) \over \bar u} \,
\int_0^{\infty} \, d \omega \, {\phi_B^{+}(\omega, \mu) \over \omega} \,,
\end{eqnarray}
where $\phi_P(u, \mu)$  is the twist-two pseudoscalar-meson LCDA.

\begin{figure}
\begin{center}
\includegraphics[width=0.45 \columnwidth]{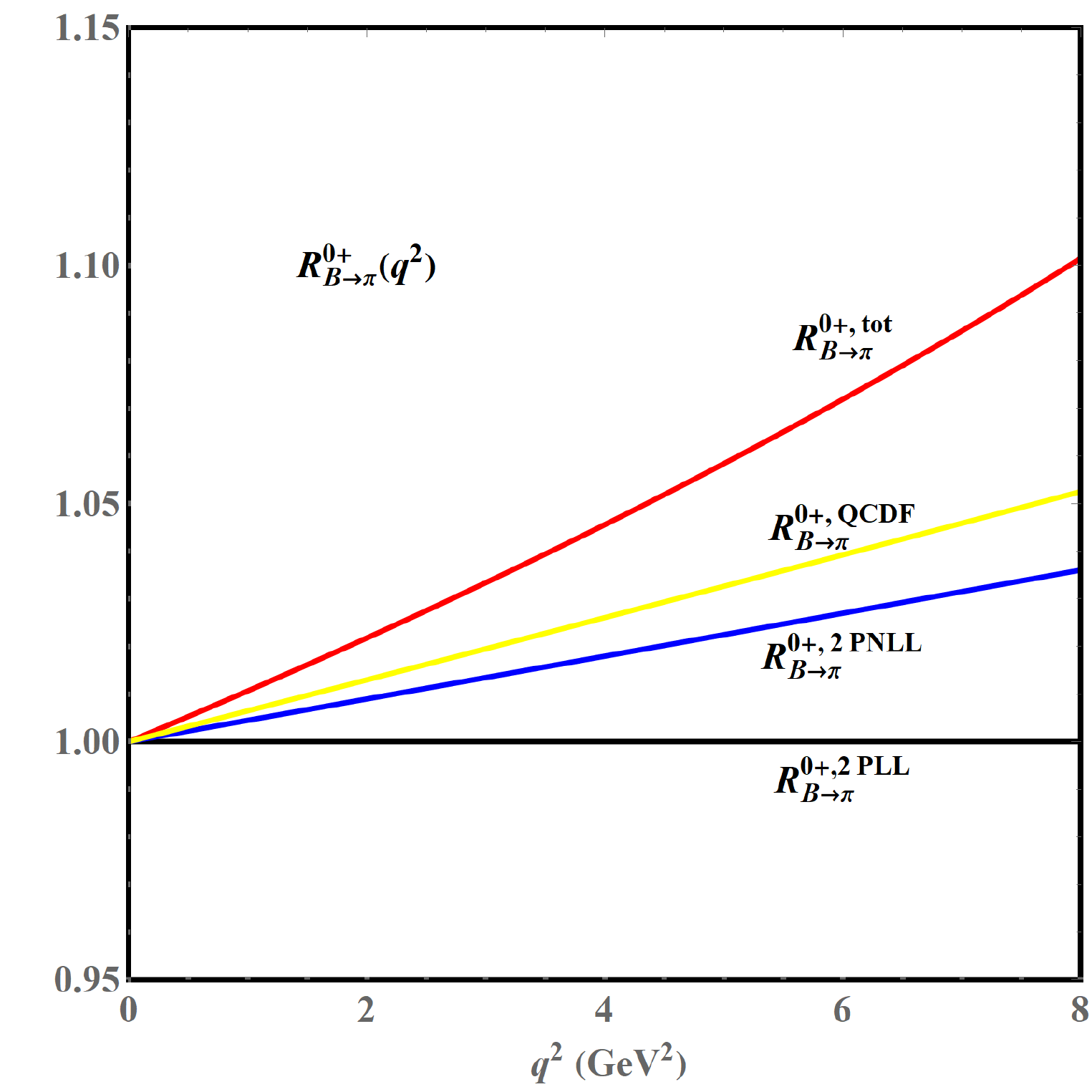}
\hspace{0.5 cm}
\includegraphics[width=0.45 \columnwidth]{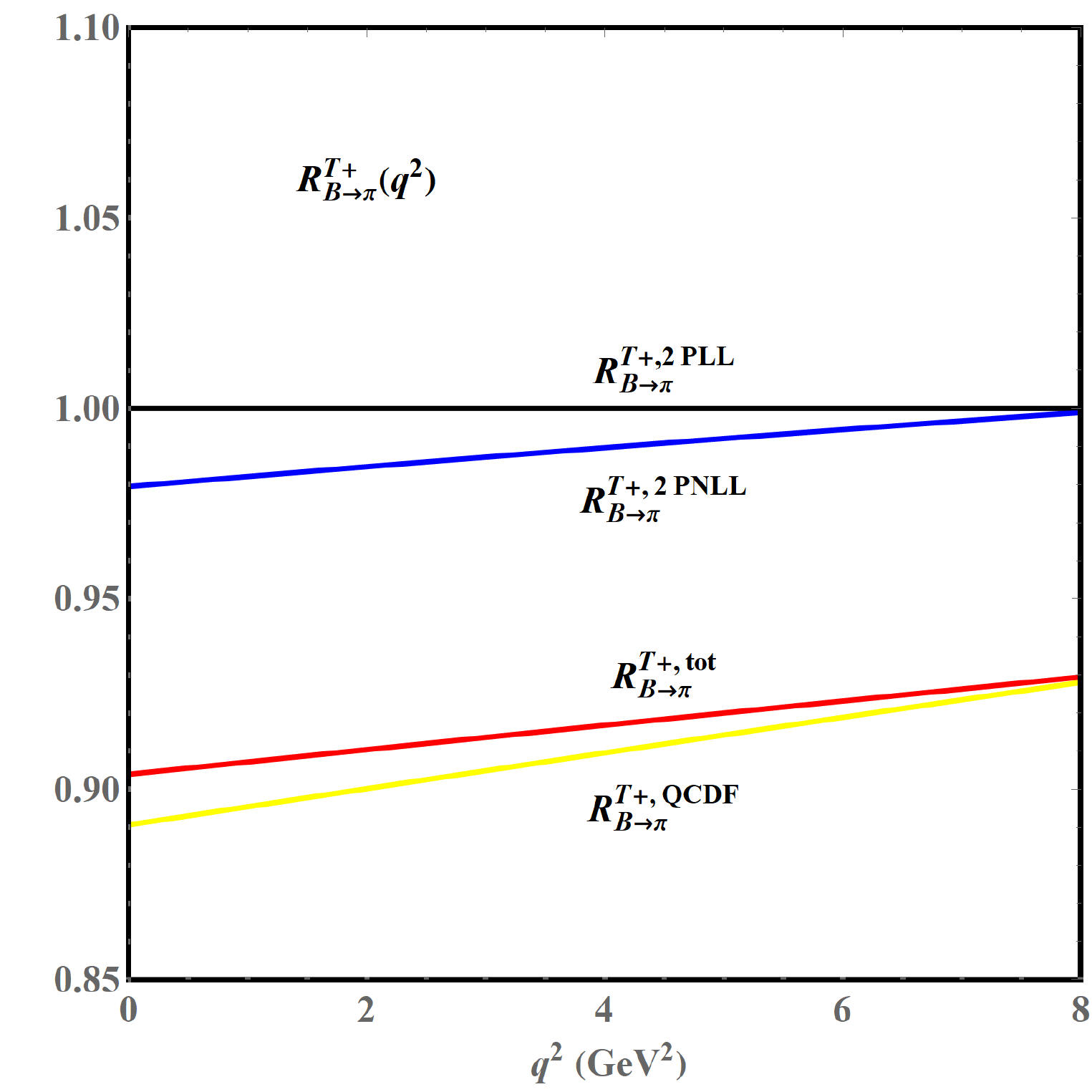}
\vspace*{0.1cm}
\caption{The large-recoil symmetry breaking effects of $B \to \pi$ form factors computed
from the LCSR approach at LL accuracy ($R_{B \to \pi}^{i \, +, \rm 2PLL}$, black),
at NLL accuracy ($R_{B \to \pi}^{i \, +, \rm 2PNLL}$, blue), and from the QCD factorization
approach ($R_{B \to \pi}^{i \, +, \rm QCDF}$, yellow). The complete LCSR predictions
for $R_{B \to \pi}^{i \, +}$ ($i=0, \, T$) with the  higher-twist $B$-meson LCDA corrections
at tree level are represented by the red curves. }
\label{fig: large-recoil symmetry breaking}
\end{center}
\end{figure}

Introducing the form-factor ratios for the semileptonic $B \to \pi$ decays
\begin{eqnarray}
R_{B \to \pi}^{0 \, +}(q^2) =  {m_B \over n \cdot p} \,
{f_{B \to \pi}^{0}(q^2) \over f_{B \to \pi}^{+}(q^2)} \,,
\qquad
R_{B \to \pi}^{T \, +}(q^2) =  {m_B \over m_B + m_{\pi}} \,
{f_{B \to \pi}^{T}(q^2) \over f_{B \to \pi}^{+}(q^2)} \,,
\end{eqnarray}
we present the theory predictions for these ratios from both our calculations and
the QCD factorization results in figure \ref{fig: large-recoil symmetry breaking}.
It is evident that both the magnitude and sign of the symmetry-breaking corrections
computed from the two QCD methods are consistent with each other and our predictions for the
large-recoil symmetry violations are generally larger than the previous LCSR
computations with pion LCDAs \cite{Ball:1998tj}.

\begin{figure}
\begin{center}
\includegraphics[width=0.45 \columnwidth]{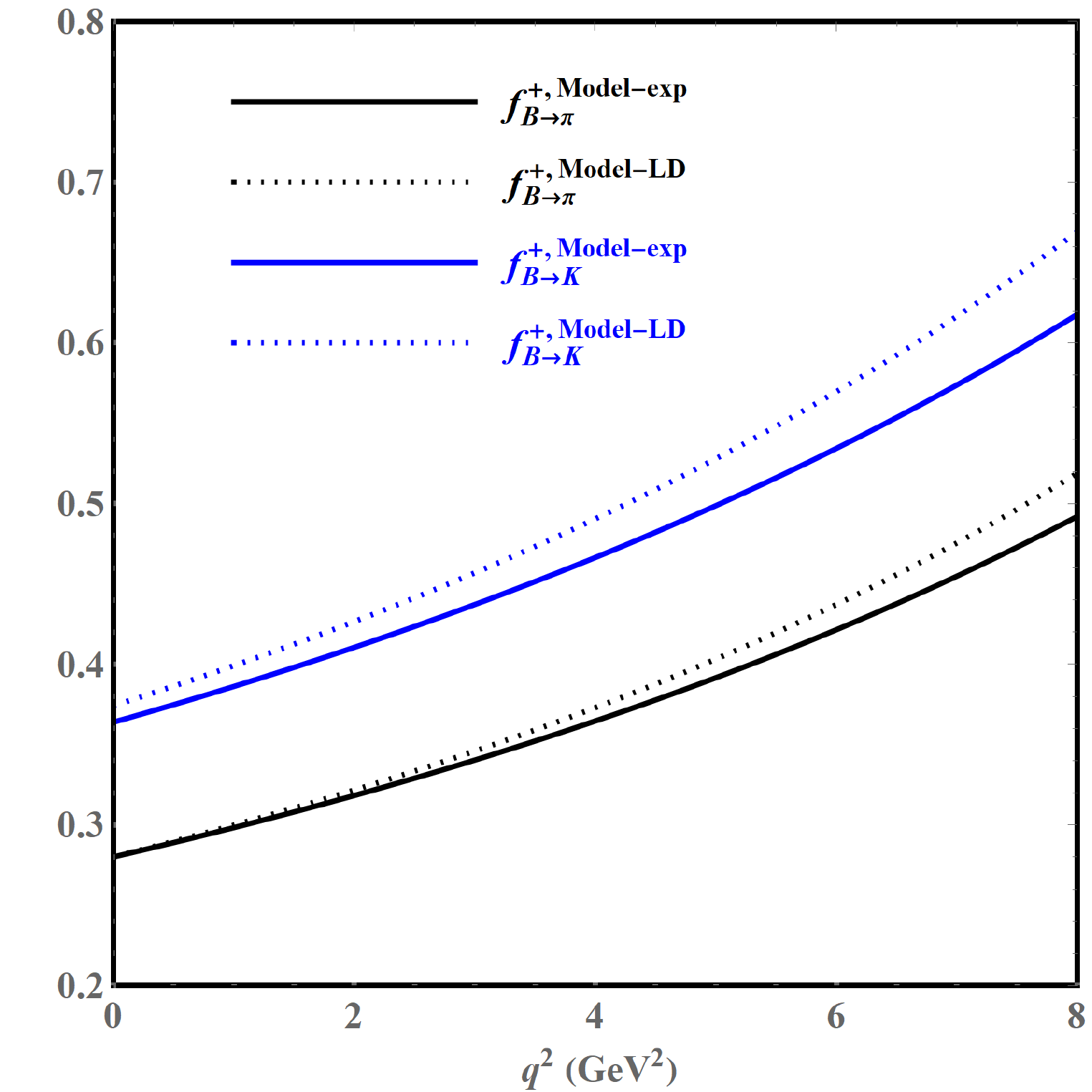}
\hspace{0.5 cm}
\includegraphics[width=0.45 \columnwidth]{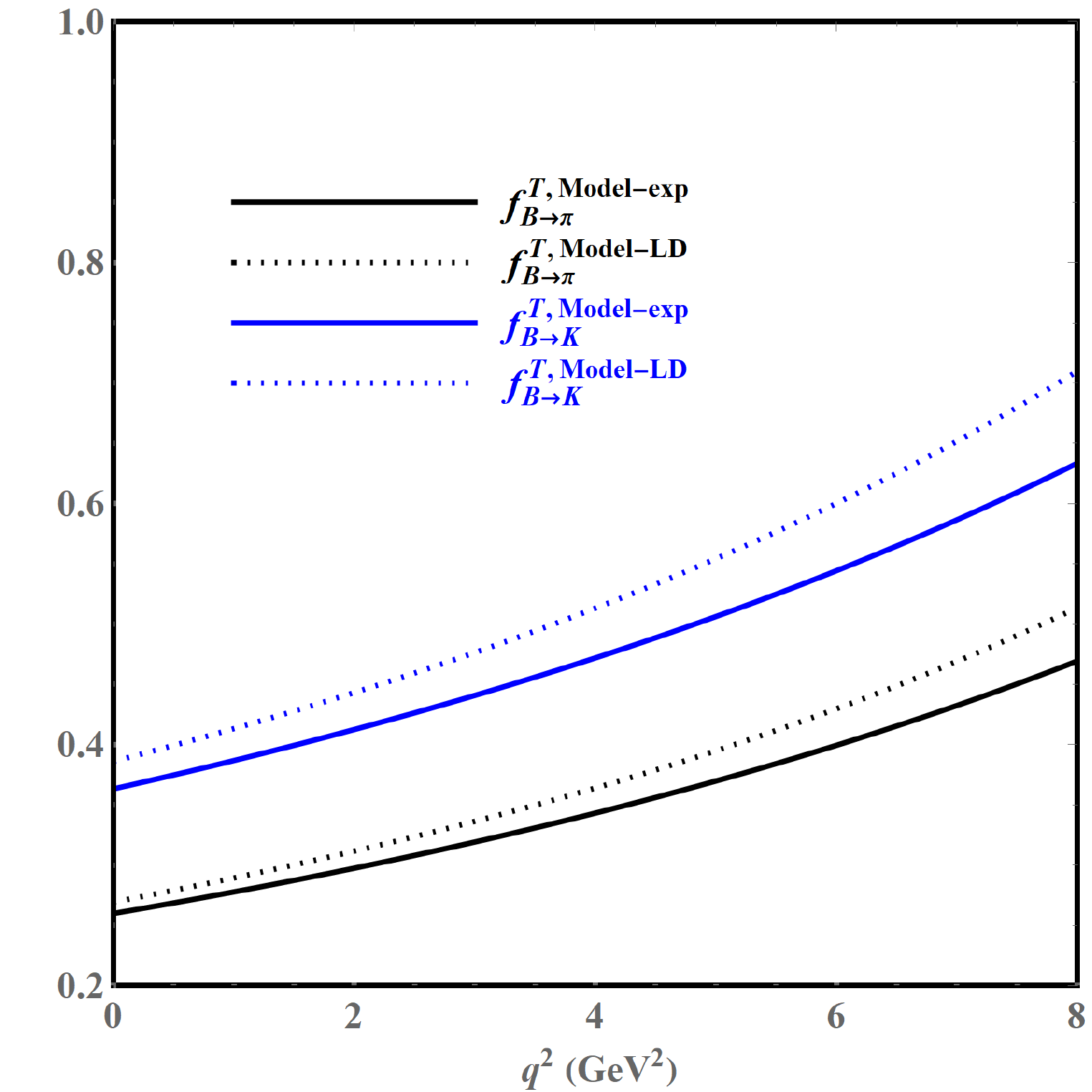}
\vspace*{0.1cm}
\caption{Dependence of the $B \to \pi, K$ form factors on the nonperturbative models
of $B$-meson LCDAs at $0 \leq q^2 \leq 8 \, {\rm GeV^2}$. The observed pattern for the
scalar form factors $f_{B \to \pi}^{0}(q^2)$ and $f_{B \to K}^{0}(q^2)$,
in analogy to  the corresponding behaviours for the vector form factors,
are not presented here for brevity. }
\label{fig: model-dependence of B to P form factors}
\end{center}
\end{figure}

To understand the model dependence of our predictions  on
the $B$-meson LCDAs, we display in figure \ref{fig: model-dependence of B to P form factors}
the obtained  $B \to \pi, K$ form factors
from both the exponential  and the local duality models as a function of the
momentum transfer $q^2$.
Taking into account the fact that the vector $B \to \pi$ form factor at $q^2=0$
has been adjusted to reproduce the values from the pion LCSR, our main prediction
is the momentum-transfer dependence of $B \to \pi, K$ form factors, which turns out to be
insensitive to the specific models of $B$-meson LCDAs
(see also \cite{Wang:2015vgv} for a similar observation).

\begin{figure}
\begin{center}
\includegraphics[width=0.40 \columnwidth]{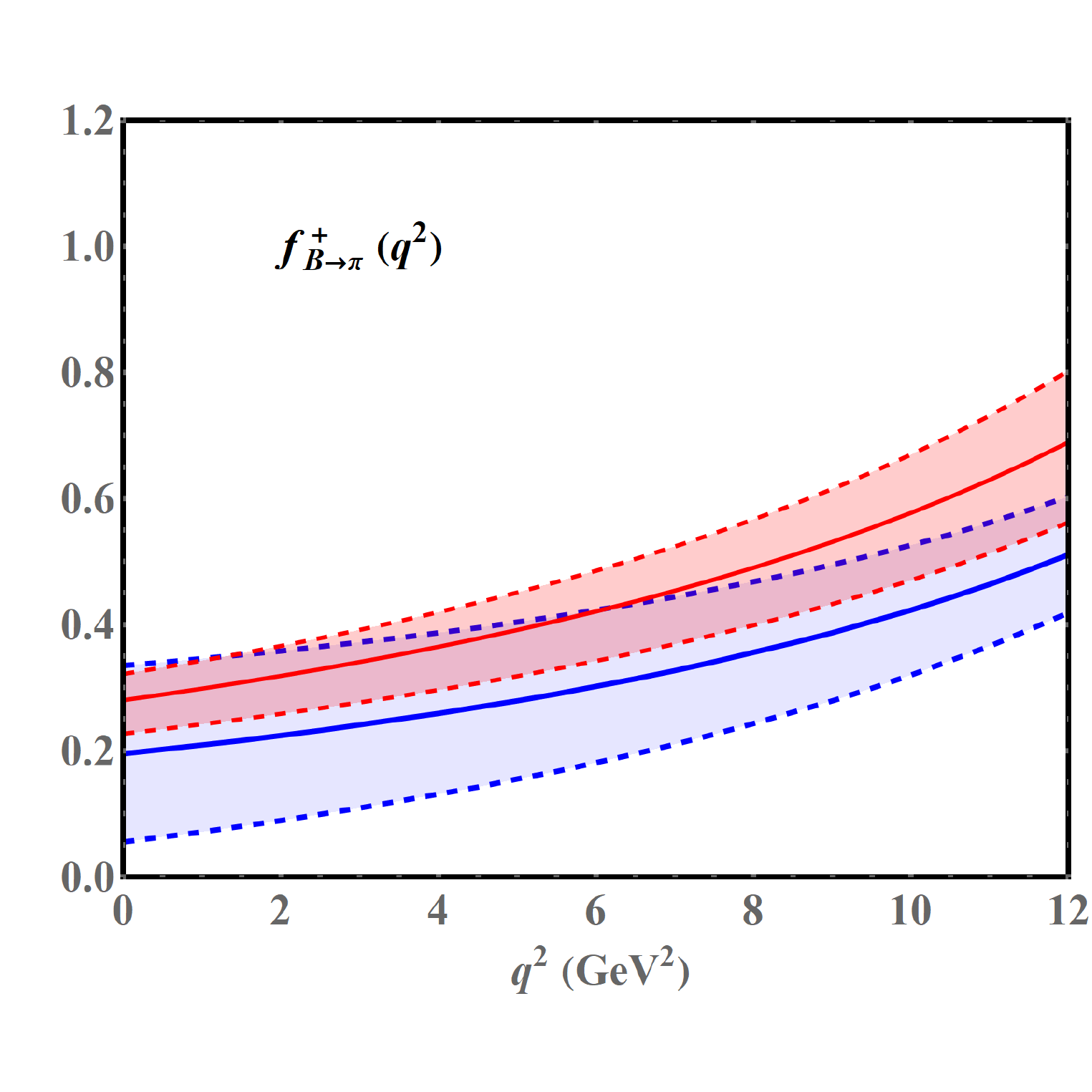}
\hspace{0.5 cm}
\includegraphics[width=0.40 \columnwidth]{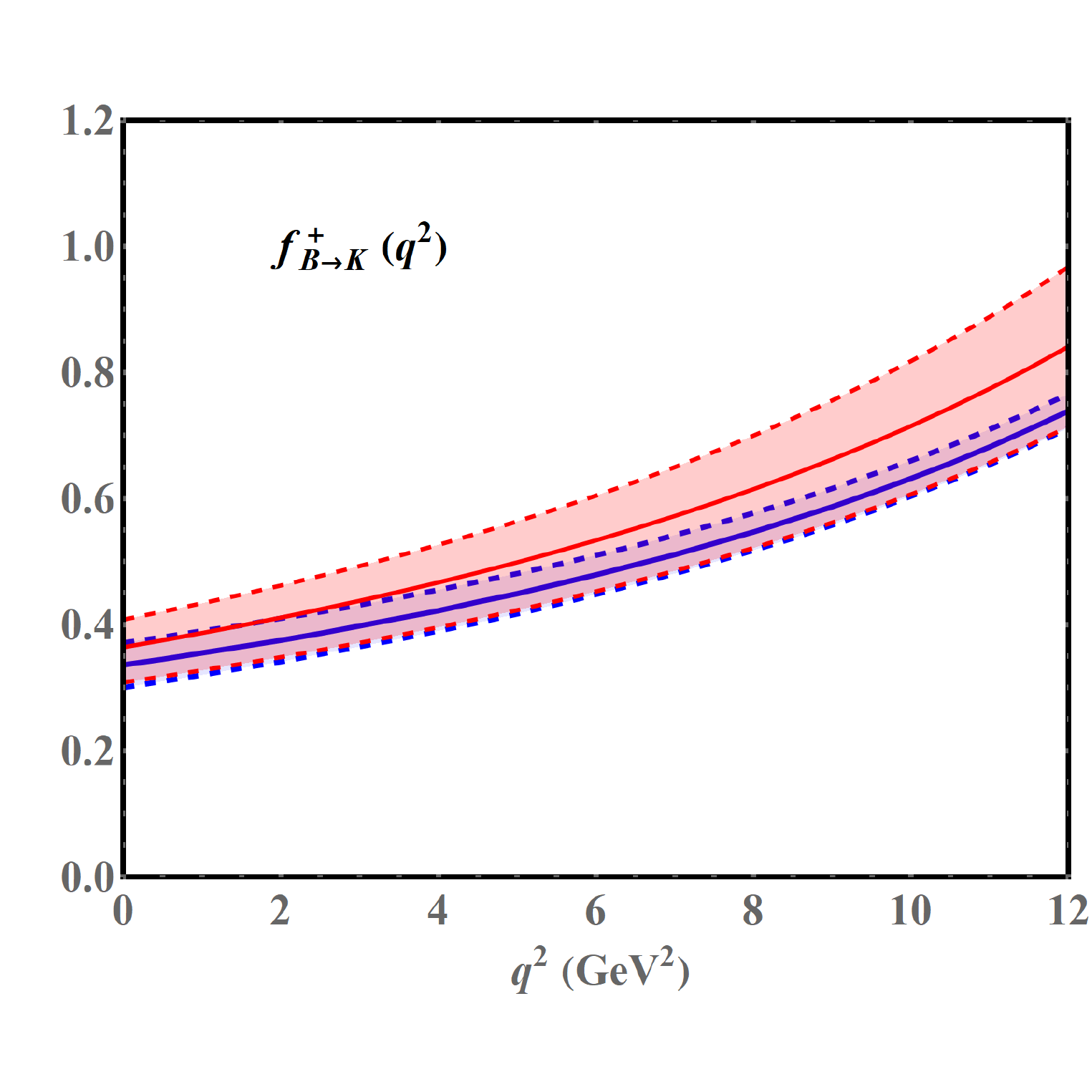}
\\
\includegraphics[width=0.40 \columnwidth]{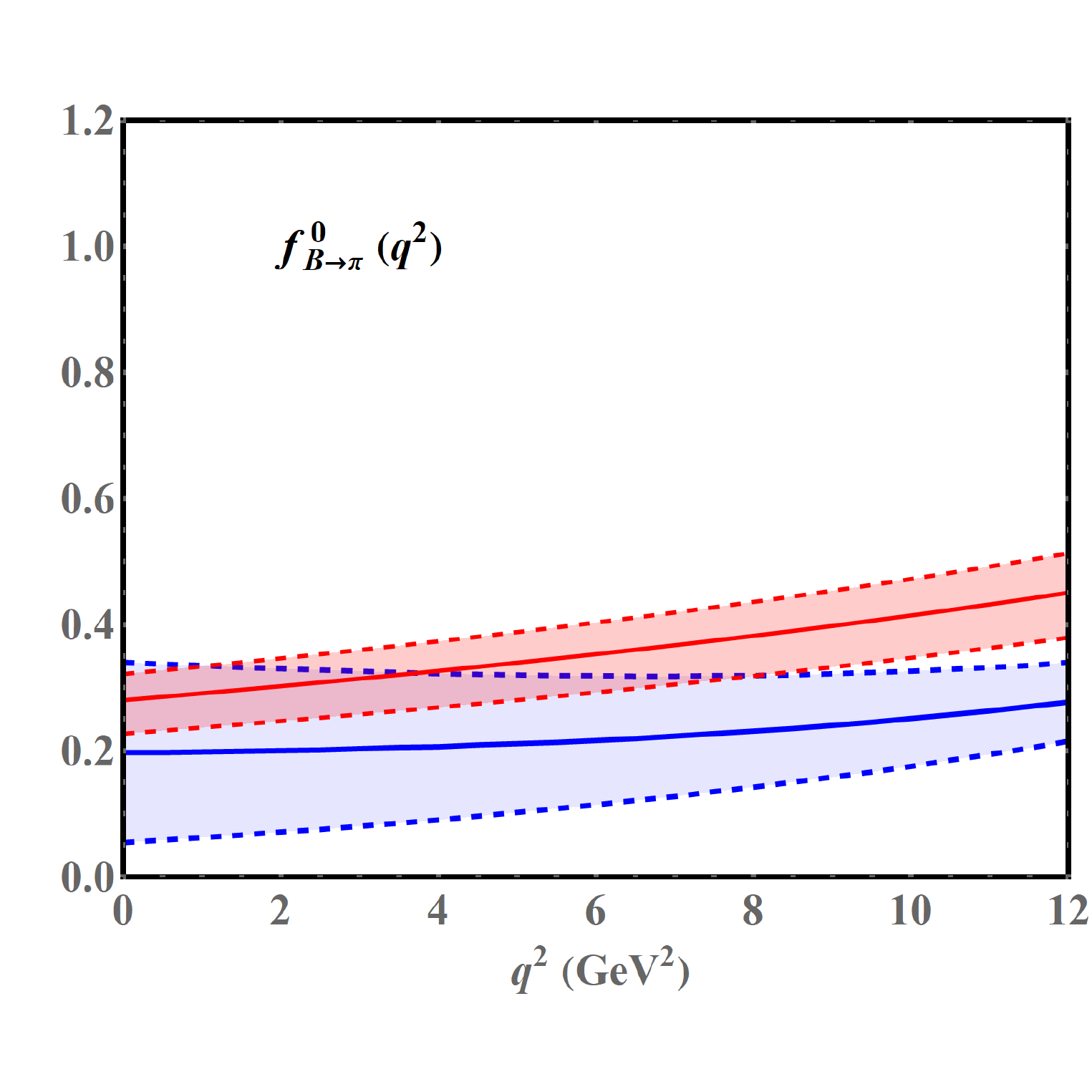}
\hspace{0.5 cm}
\includegraphics[width=0.40 \columnwidth]{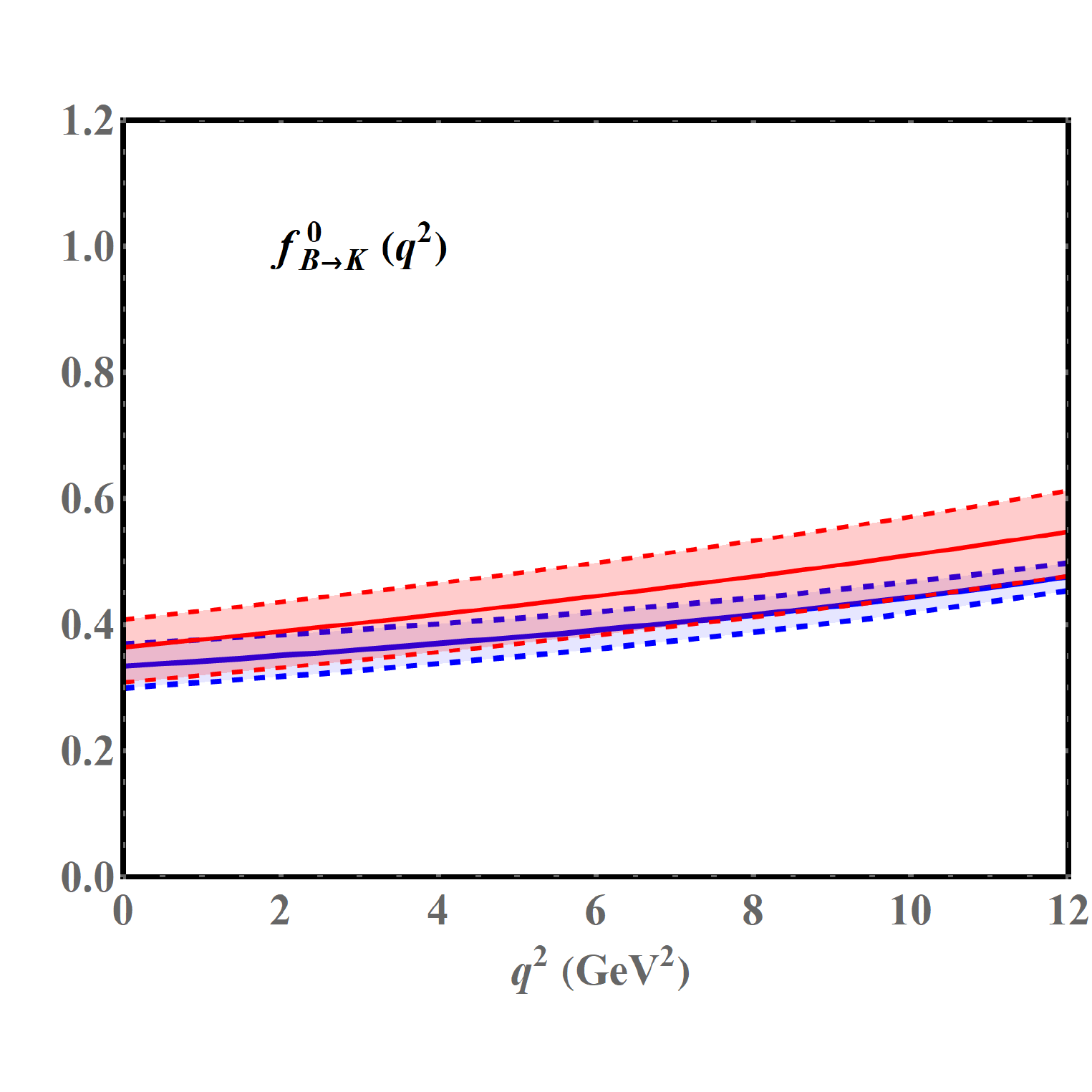}
\\
\includegraphics[width=0.40 \columnwidth]{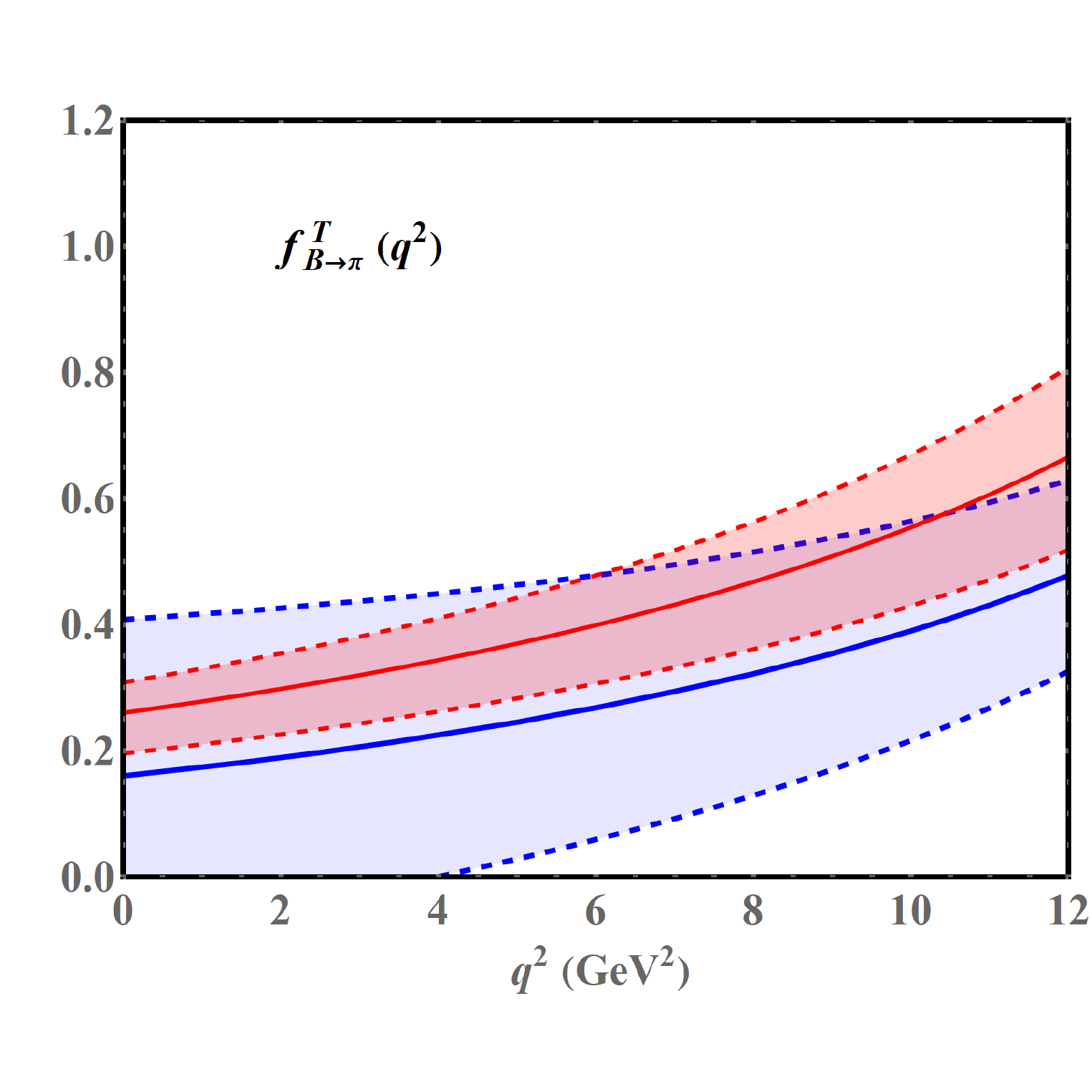}
\hspace{0.5 cm}
\includegraphics[width=0.40 \columnwidth]{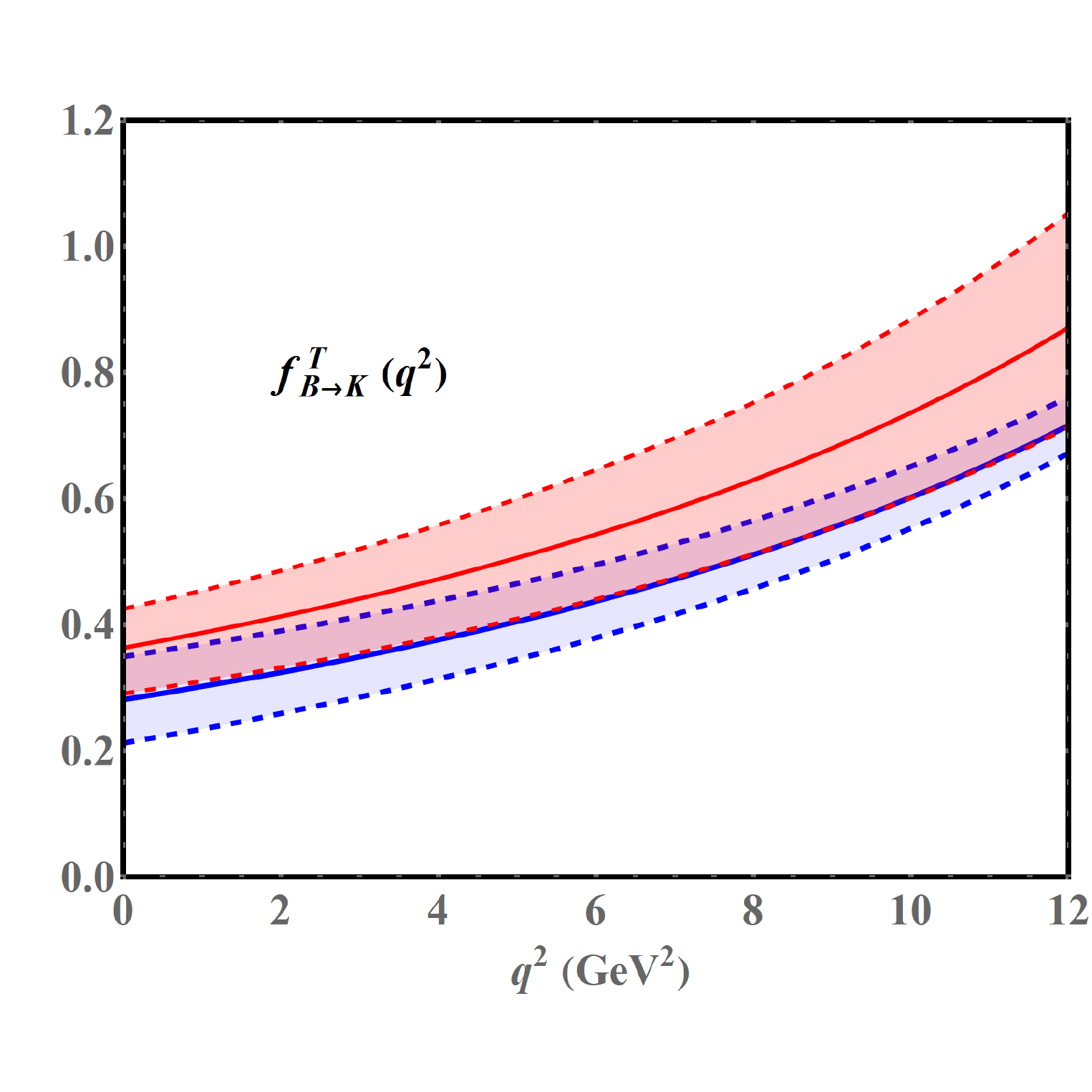}
\vspace*{0.1cm}
\caption{The momentum-transfer dependence of $B \to \pi, K$ form factors predicted
from the $B$-meson LCSR are displayed with the pink bands.
For a comparison, we also present the Lattice QCD calculations from
Fermilab/MILC Collaborations \cite{Lattice:2015tia,Bailey:2015nbd,Bailey:2015dka}
with an extrapolation to small $q^2$ in terms of the $z$-series
expansion (\ref{z expansion of B to P FFs})  as indicated by the blue bands.}
\label{fig: q2-dependence of B to P form factors from LCSR}
\end{center}
\end{figure}

\begin{table}[t!bph]
\begin{center}
\begin{tabular}{|c|c|c|c|c|c|c|c|c|}
  \hline
  \hline
  &  &  &  &  &  &  &  &  \\
  Parameters & Central \,\, value & $\lambda_B$ & $\sigma_1$ & $\mu$ & $\nu$ & $M^2$ & $s_0$ & $\phi_B^{\pm}(\omega)$ \\
  &  &  &  &  &  &  &  &  \\
  \hline
   &  &  &  &  &  &  &  &  \\
  $f_{B \to \pi}^{+, 0}(0)$ & 0.280 & $^{-0.030}_{+0.031}$ & $^{-0.012}_{+0.013}$ & $^{+0.000}_{-0.032}$ & - & $^{+0.012}_{-0.017}$ & $^{+0.014}_{-0.014}$ & - \\
  &  &  &  &  &  &  &  &  \\
  $b_{1, \pi}^{+}$ & $-2.77$ & $^{+0.05}_{-0.02}$ & $^{+0.02}_{-0.01}$ & $^{+0.09}_{-0.16}$ & - & $^{+0.02}_{-0.03}$ & $^{+0.07}_{-0.07}$ & $^{+0.00}_{-0.64}$ \\
  &  &  &  &  &  &  &  &  \\
  $b_{1, \pi}^{0}$ & $-4.88$ & $^{-0.10}_{+0.11}$ & $^{-0.04}_{+0.04}$ & $^{+0.17}_{-0.61}$ & - & $^{+0.04}_{-0.06}$ & $^{+0.11}_{-0.11}$ & $^{+0.00}_{-0.37}$ \\
  &  &  &  &  &  &  &  &  \\
  \hline
   &  &  &  &  &  &  &  &  \\
  $f_{B \to \pi}^{T}(0)$ & 0.260 & $^{-0.031}_{+0.031}$ & $^{-0.013}_{+0.013}$ & $^{+0.000}_{-0.044}$ & $^{-0.017}_{+0.025}$ & $^{+0.011}_{-0.016}$ & $^{+0.013}_{-0.014}$ & - \\
  &  &  &  &  &  &  &  &  \\
  $b_{1, \pi}^{T}$ & $-3.14$ & $^{+0.05}_{-0.02}$ & $^{+0.02}_{-0.01}$ & $^{+0.21}_{-0.57}$ & $^{+0.05}_{-0.06}$ & $^{+0.02}_{-0.03}$ & $^{+0.07}_{-0.07}$ & $^{+0.00}_{-0.67}$ \\
   &  &  &  &  &  &  &  &  \\
  \hline
   &  &  &  &  &  &  &  &  \\
  $f_{B \to K}^{+, 0}(0)$ & 0.364 & $^{-0.035}_{+0.034}$ & $^{-0.014}_{+0.014}$ & $^{+0.000}_{-0.032}$ & - & $^{+0.010}_{-0.014}$ & $^{+0.008}_{-0.009}$ & - \\
  &  &  &  &  &  &  &  &  \\
  $b_{1, K}^{+}$ & $-3.04$ & $^{+0.02}_{-0.00}$ & $^{+0.00}_{-0.00}$ & $^{+0.04}_{-0.06}$ & - & $^{+0.05}_{-0.07}$ & $^{+0.05}_{-0.06}$ & $^{+0.00}_{-0.76}$ \\
  &  &  &  &  &  &  &  &  \\
  $b_{1, K}^{0}$ & $-4.56$  & $^{-0.13}_{+0.14}$ & $^{-0.06}_{+0.06}$ & $^{+0.08}_{-0.41}$ & - & $^{+0.07}_{-0.10}$ & $^{+0.07}_{-0.08}$ & $^{+0.00}_{-0.42}$ \\
  &  &  &  &  &  &  &  &  \\
  \hline
  &  &  &  &  &  &  &  &  \\
  $f_{B \to K}^{T}(0)$ & 0.363 & $^{-0.038}_{+0.038}$  & $^{-0.016}_{+0.016}$ & $^{+0.000}_{-0.048}$ & $^{-0.022}_{+0.033}$ & $^{+0.011}_{-0.014}$ & $^{+0.009}_{-0.009}$ & $^{+0.023}_{-0.000}$ \\
  &  &  &  &  &  &  &  &  \\
  $b_{1, K}^{T}$ & $-3.47$ & $^{+0.02}_{-0.00}$ & $^{-0.00}_{+0.01}$ & $^{+0.16}_{-0.46}$ & $^{+0.05}_{-0.06}$ & $^{+0.05}_{-0.07}$ & $^{+0.05}_{-0.06}$ & $^{+0.00}_{-0.79}$ \\
  &  &  &  &  &  &  &  &  \\
  \hline
  \hline
\end{tabular}
\end{center}
\caption{Theory predictions for the shape parameters and the normalizations of $B \to \pi, K$ form factors at $q^2=0$
entering the $z$ expansion (\ref{z expansion of B to P FFs}) with the dominant uncertainties
from variations of different input parameters.}
\label{table: fitted results for the shape parameters}
\end{table}

As already discussed in \cite{Khodjamirian:2006st,Wang:2015vgv}, the light-cone operator product expansion (OPE) of the
vacuum-to-$B$-meson correlation function (\ref{correlator: definition}) can  be verified only at large
hadronic recoil. We will extrapolate the LCSR predictions of $B \to \pi, K$ form factors at $q^2 \leq 8 \, {\rm GeV}^2$
to the full kinematic region by applying the $z$-series expansion,
where the entire cut $q^2$-plane is mapped onto the unit disk $|z(q^2, \, t_0)|\leq 1$
with the conformal transformation
\begin{eqnarray}
z(q^2, t_0) = \frac{\sqrt{t_{+}-q^2}-\sqrt{t_{+}-t_0}}
{\sqrt{t_{+}-q^2}+\sqrt{t_{+}-t_0}}  \,.
\end{eqnarray}
Here, $t_{+} = (m_B + m_{P})^2$ is determined by the threshold of the lowest continuum state
which can be generated by the weak transition currents in QCD.
The auxiliary parameter $t_0$ determining the $q^2$ point to be mapped onto the origin
of the complex $z$ plane will be further taken as \cite{Khodjamirian:2017fxg}
\begin{eqnarray}
t_0 = (m_B + m_P) \, (\sqrt{m_B} + \sqrt{m_P})^2 \,.
\end{eqnarray}
For  concreteness, we will adopt the simplified series expansion for $B \to P$ form factors
originally proposed in \cite{Bourrely:2008za} (see \cite{Boyd:1994tt} for an alternative parametrization)
\begin{eqnarray}
f_{B \to P}^{+, T}(q^2) &=&  {f_{B \to P}^{+, T}(0) \over 1 - q^2/m_{B_{(s)}^{\ast}}^2} \,
\bigg \{ 1 + \, \sum_{k=1}^{N-1}   \, b_{k, P}^{+, T}  \,
\bigg  ( z(q^2, \, t_0)^k -  z(0, \, t_0)^k  \nonumber \\
&& - \, (-1)^{N-k} \, {k \over N} \,
\left [  z(q^2, \, t_0)^N -  z(0, \, t_0)^N \right ]  \bigg  ) \bigg \}   \,,
\nonumber \\
f_{B \to P}^{0}(q^2) &=&  f_{B \to P}^{0}(0) \,
\left \{ 1 +  \, \sum_{k=1}^{N}   \, b_{k, P}^{0}  \,
\left (  z(q^2, \, t_0)^k -  z(0, \, t_0)^k  \right )   \right \}   \,,
\label{z expansion of B to P FFs}
\end{eqnarray}
where the threshold behaviour at $q^2=t_{+}$ has been implemented and
we will truncate the $z$-series at $N=2$  for the vector (tensor) form factors
and at $N=1$ for the scalar form factors.

Matching the $B$-meson LCSR calculations in the kinematic region $- 6 \, {\rm GeV^2} \leq q^2 \leq 8 \, {\rm GeV^2}$
with the $z$-series expansion (\ref{z expansion of B to P FFs}) leads to our main predictions for the
$q^2$ dependence of $B \to \pi, K$ form factors displayed
in figure \ref{fig: q2-dependence of B to P form factors from LCSR}, where the theory uncertainties due to varying
different input parameters discussed in Section \ref{section: theory inputs} are also included.
We further display the low $q^2$-extrapolation of the  Lattice QCD  predictions from Fermilab/MILC Collaborations
\cite{Lattice:2015tia,Bailey:2015nbd,Bailey:2015dka}  in the same figure for a comparison.
While we find a fair agreement of the two calculations, our predictions for the three $B \to \pi$ form factors
are more precise than the corresponding  Lattice QCD  results.
The resulting shape parameters and the normalizations of $B \to \pi, K$ form factors at $q^2=0$ entering
the $z$-series (\ref{z expansion of B to P FFs}) are collected in Table \ref{table: fitted results for the shape parameters}
with the numerically important uncertainties.
We can readily observe that the dominant theory uncertainties for the form factors at $q^2=0$ originate from the variations
of the inverse moment $\lambda_B(\mu_0)$, while the model dependence of the $B$-meson LCDAs leads to the most significant
errors for the shape parameters $b_{1, P}^{i}$ ($i=+, \, 0, \, T$).
In particular,  the tensor $B \to \pi, K$ form factors appear to suffer from larger uncertainties compared with the
corresponding vector and scalar form factors, due to the sizeable errors from  variations of the QCD renormalization scale
of the tensor current.

\subsection{Phenomenological aspects of $B \to \pi l \nu$}

Having at our disposal the theory predictions for $B \to \pi$ form factors,
we proceed to explore  phenomenological aspects of the semileptonic $B \to \pi \ell \nu$  decays,
which serves as the golden channel for the determination of CKM matrix element $|V_{ub}|$ exclusively
(see \cite{Kou:2018nap} for the future advances of precision measurements of Belle II).
It is straightforward to write down the differential decay rate for $B \to \pi \ell \nu$
\begin{eqnarray}
{d \,  \Gamma (B \to \pi \ell \nu)\over d q^2}
&=& {G_F^2 \, |V_{ub}|^2 \over 192 \, \pi^3 \, m_B^3} \, \lambda^{3/2}(m_B^2, m_{\pi}^2, q^2) \,
\left (  1 -{m_l^2 \over  q^2} \right )^2  \,  \left (  1 + {m_l^2 \over  2 \, q^2} \right ) \,
\bigg [ |f_{B \to \pi}^{+} (q^2)|^2  \nonumber \\
&& +  \, {3 \, m_l^2 \, (m_B^2 - m_{\pi}^2)^2 \over \lambda(m_B^2, m_{\pi}^2, q^2) \, (m_l^2 + 2\, q^2) } \,
|f_{B \to \pi}^{0} (q^2)|^2   \bigg ]  \,,
\end{eqnarray}
where $\lambda(a, b, c) = a^2 + b^2 + c^2- 2 \, ab -2 \, ac - 2 \, bc$.

\begin{figure}
\begin{center}
\includegraphics[width=0.65 \columnwidth]{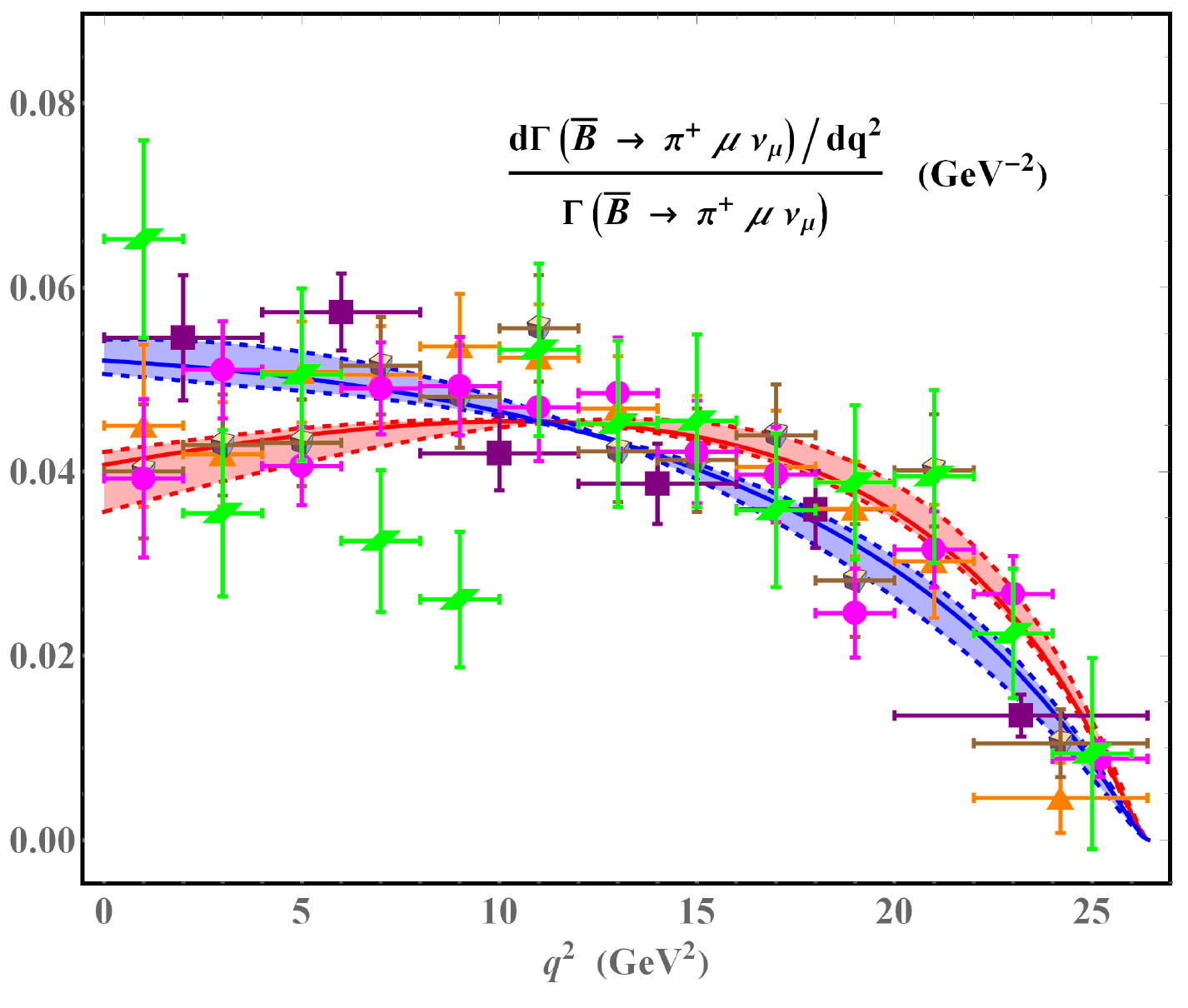}
\hspace{0.5 cm}
\includegraphics[width=0.65 \columnwidth]{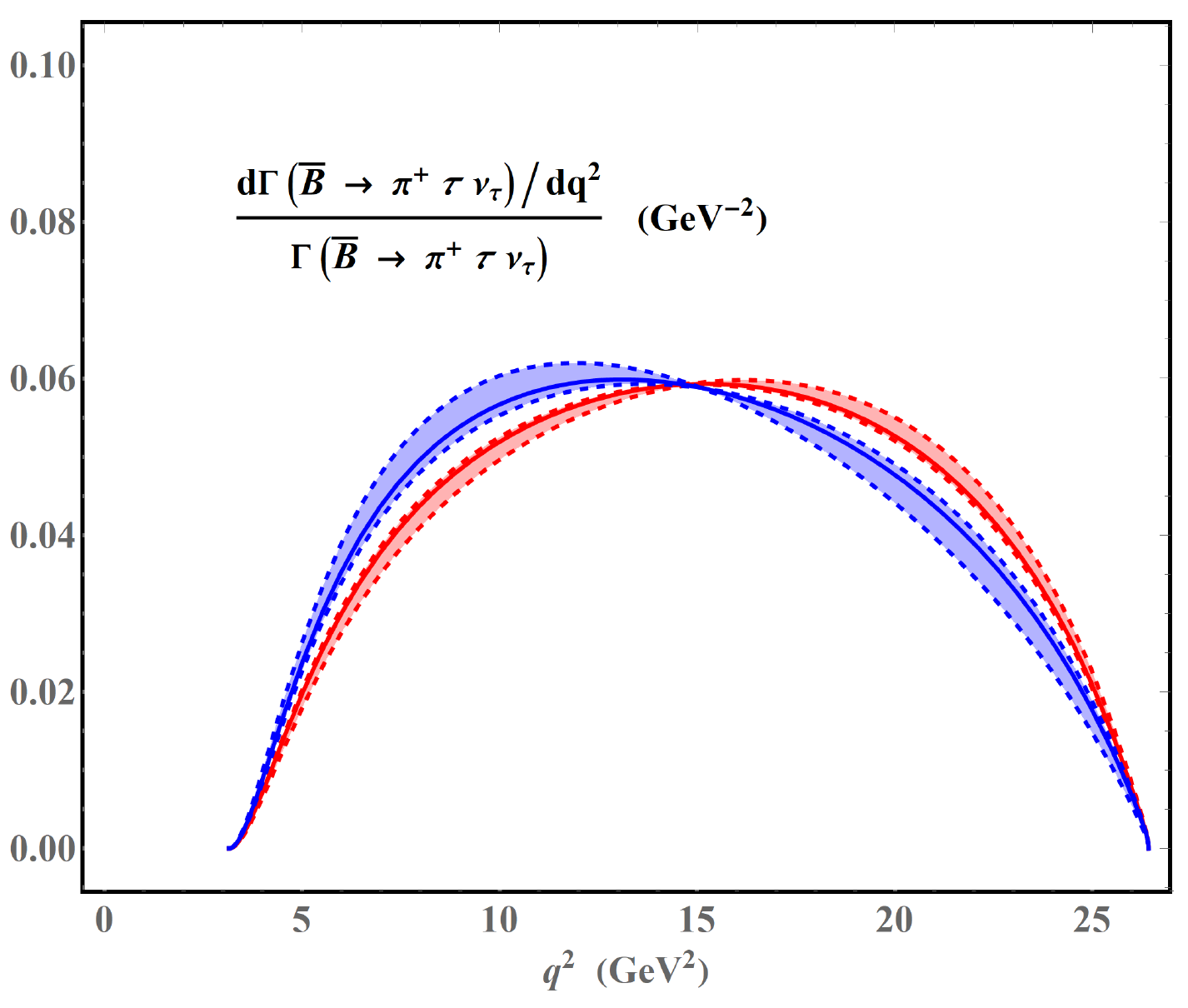}
\vspace*{0.1cm}
\caption{The normalized differential $q^2$ distributions of the semileptonic
$B \to \pi \ell \nu$ ($\ell=\mu\,, \tau$)  decays with the form factors computed from the $B$-meson LCSR
with an extrapolation to the whole  kinematical  region (pink bands) and from
the pion LCSR with the $z$-series expansion (blue bands).
We also present the experimental data points $B \to \pi \mu \nu_{\mu}$ 
from \cite{delAmoSanchez:2010af} (purple squares),
\cite{delAmoSanchez:2010zd} (orange triangles), \cite{Lees:2012vv} (brown hexahedrons),
\cite{Ha:2010rf} (magenta circles) and \cite{Sibidanov:2013rkk} (green parallelograms).}
\label{fig:normalized differential q2 distributions of B to pi l nu}
\end{center}
\end{figure}

Following the strategy presented in \cite{Khodjamirian:2011ub}, the extraction of $|V_{ub}|$
can be achieved by introducing the following quantity
\begin{eqnarray}
\Delta \zeta_{\ell}(q_1^2, q_2^2) = {1 \over |V_{ub}|^2}  \,
\int_{q_1^2}^{q_2^2} \, dq^2 \,   {d \, \Gamma (B \to \pi \ell \nu)  \over d q^2}  \, \,.
\end{eqnarray}
Employing the predicted $B \to \pi$ form factor $f_{B \to \pi}^{+}(q^2)$ from
the $B$-meson LCSR with an extrapolation toward large $q^2$, we obtain
\begin{eqnarray}
\Delta \zeta_{\mu}(0, 12 \, {\rm GeV^2}) &=&
\left ( 5.17\,{}^{+1.23}_{-1.07} \, \big|_{\lambda_B}\,{}^{+0.50}_{-0.44} \, \big|_{\sigma_1}
\,{}^{+0.44}_{-0.59} \, \big|_{M^2}
\,{}^{+0.49}_{-0.47} \, \big|_{s_0}
\,{}^{+0.38}_{-0.00} \, \big|_{\phi_B^{\pm}} \right ) \,\,\, {\rm ps}^{-1} \nonumber   \\
&=& 5.17^{+1.65}_{-1.85} \,\,\, {\rm ps}^{-1} \,,
\label{zeta integral}
\end{eqnarray}
where the negligibly small uncertainties due to  variations of the remaining parameters
(not explicitly displayed here) are also taken into account in the total uncertainty.
Taking advantage of the experimental measurements
of $B \to \pi \mu \nu_{\mu}$  from BaBar and Belle Collaborations \cite{Lees:2012vv,Sibidanov:2013rkk},
the CKM matrix element $|V_{ub}|$ is determined as
\begin{eqnarray}
|V_{ub}| = \bigg ( 3.23\,{}^{+0.66}_{-0.48} \big |_{\rm th.}\,{}^{+0.11}_{-0.11} \big |_{\rm exp.} \bigg )
\times 10^{-3} \,,
\end{eqnarray}
which are in agreement with the averaged exclusive determinations presented in PDG \cite{Tanabashi:2018oca}
and the previous LCSR calculations \cite{Khodjamirian:2011ub,Wang:2015vgv}, but are significantly lower than the
averaged inclusive determinations reported in \cite{Tanabashi:2018oca}
\begin{eqnarray}
|V_{ub}|_{\rm inc.} = \bigg ( 4.49 \pm 0.15 \,{}^{+0.16}_{-0.17}  \pm 0.17 \bigg )
\times 10^{-3} \,.
\end{eqnarray}

We further display in figure \ref{fig:normalized differential q2 distributions of B to pi l nu}
our predictions for the normalized differential $q^2$ distributions of
$B \to \pi \ell \nu$ ($\ell=\mu\,, \tau$) in the whole kinematical  region,
where the experimental measurements from  BaBar and Belle Collaborations are also displayed for a comparison.
On account of the substantial cancellation of theory uncertainties between the differential and the total decay rates
of $B \to \pi \ell \nu$ , the momentum-transfer dependence of the normalized differential distributions
suffers from much less uncertainty than the semileptonic
$B \to \pi$ form factors shown in figure \ref{fig: q2-dependence of B to P form factors from LCSR}.
The future precision measurements of $B \to \pi \ell \nu$ from Belle II
(with remarkably high accuracy of ${\cal O}(1.4 \, \%)$ \cite{Kou:2018nap}) will be helpful to distinguish the theory
predictions based upon the distinct LCSR  methods presented in
figure \ref{fig:normalized differential q2 distributions of B to pi l nu}.

\subsection{Phenomenological aspects of $B \to K \nu \nu$}

\begin{figure}
\begin{center}
\includegraphics[width=0.65 \columnwidth]{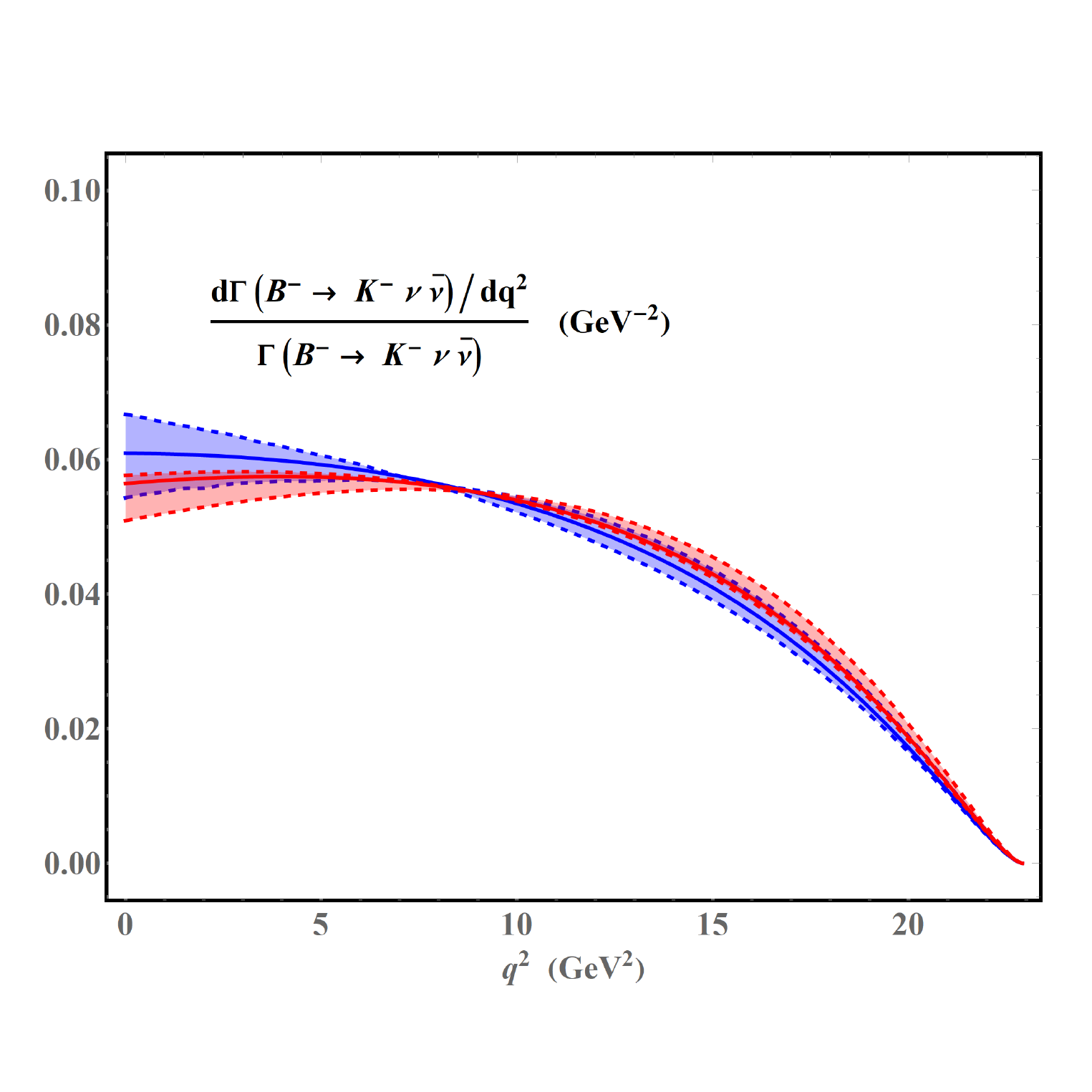}
\vspace*{0.1cm}
\caption{The normalized differential $q^2$ distribution of $B \to K \nu \nu$
computed with the form factors from the $B$-meson LCSR (pink band) and the Lattice
QCD simulations \cite{Bailey:2015dka} (blue band).}
\label{fig:normalized differential q2 distributions of B to K nu nu}
\end{center}
\end{figure}

The information of $B \to K$ form factors enables us to investigate the  rare exclusive $B \to K \nu \nu$
decays induced by the flavour-changing neutral current $b \to s \nu \nu$.
An important advantage of such process over the more complicated
$B \to K^{(\ast)} \ell \ell$ decays \cite{Beneke:2001at,Khodjamirian:2010vf,Khodjamirian:2012rm}
lies in the fact that  the strong interaction dynamics of $B \to K \nu \nu$ is completely encoded in
the semileptonic $B$-meson form factors.
It is straightforward to write down the differential decay rate for $B \to K \nu \nu$
\begin{eqnarray}
{d \, \Gamma(B \to K \nu \nu) \over d q^2} &=&  {G_F^2 \, \alpha_{\rm em}^2  \over 256 \, \pi^5}  ¡¢£¬
\frac{\lambda^{3/2}(m_B^2, m_K^2, q^2)}{m_B^2\, \sin^4 \, \theta_W} \, |V_{tb} \, V_{ts}^{\ast}|^2  \,
\left [ X_t \left ({m_t^2 \over m_W^2}, \, {m_H^2 \over m_t^2}, \, \sin \, \theta_W, \, \mu \right ) \right ]^2 \, \nonumber \\
&& \times \, |f_{B \to K}^{+}(q^2)|^2 \,,
\label{differential decay width of B to K nu nu}
\end{eqnarray}
where the short-distance Wilson coefficient $X_t$ has been computed at NLO in QCD \cite{Buchalla:1998ba,Buchalla:1992zm,Misiak:1999yg}
and at two loops in the electroweak Standard Model (SM) \cite{Brod:2010hi}.
For the numerical analysis,  the intervals of various electroweak parameters
entering (\ref{differential decay width of B to K nu nu}) will be taken from \cite{Brod:2010hi}.
Our prediction for the normalized differential $q^2$ distribution of $B \to K \nu \nu$
with the vector $B \to K$ form factor computed from the $B$-meson LCSR is presented
in figure \ref{fig:normalized differential q2 distributions of B to K nu nu},
where the theory results with the Lattice QCD form factor \cite{Bailey:2015dka}
are also displayed.

\begin{table}[t!bph]
\begin{center}
\begin{tabular}{|c|c|c|}
  \hline
  \hline
  &   &    \\
  $[q_1^2, \, q_2^2]$  \,\, (${\rm in \,\, GeV}^2$) & $10^6 \times \Delta {\cal BR}(q_1^2, q_2^2)$ & $10^2 \times R_{K \pi}(q_1^2, q_2^2)$ \\
  &   &    \\
  \hline
  &   &    \\
  $[0.0, 1.0]$ & $0.34^{+0.09}_{-0.10}$  &  $5.33^{+0.60}_{-0.39}$  \\
  &   &    \\
  $[1.0, 2.5]$ & $0.52^{+0.13}_{-0.15}$  &  $5.27^{+0.59}_{-0.39}$  \\
  &   &    \\
  $[2.5, 4.0]$ & $0.52^{+0.14}_{-0.15}$  &  $5.18^{+0.57}_{-0.38}$  \\
  &   &    \\
  $[4.0, 6.0]$ & $0.69^{+0.19}_{-0.20}$  &  $5.06^{+0.55}_{-0.38}$  \\
  &   &    \\
  $[6.0, 8.0]$ & $0.68^{+0.19}_{-0.20}$  &  $4.90^{+0.53}_{-0.37}$  \\
  &   &    \\
  $[0.0, 8.0]$ & $2.75^{+0.64}_{-0.81}$  &  $5.11^{+0.56}_{-0.38}$  \\
  &   &    \\
  $[0, (m_B-m_K)^2]$ & $6.02^{+1.68}_{-1.76}$  &  $4.06^{+0.39}_{-0.30}$  \\
  &   &    \\
  \hline
  \hline
\end{tabular}
\end{center}
\caption{Theory predictions for the partial branching fractions of $B \to K \nu \nu$
and the binned distributions of the precision observable $R_{K \pi}(q_1^2, q_2^2)$
with $B \to \pi,  K$ form factors computed from the $B$-meson LCSR and the
$z$-series expansion. }
\label{table: binned distributions of B to K nu nu}
\end{table}

To facilitate the comparison with the future Belle II data, we further
introduce the partial branching fraction of $B \to K \nu \nu$
\begin{eqnarray}
\Delta {\cal BR}(q_1^2, q_2^2) = \tau_{B^0} \,  \int_{q_1^2}^{q_2^2} \, dq^2 \,\,
{d \, \Gamma(B \to K \nu \nu) \over d q^2}\,,
\end{eqnarray}
whose predictions for the selected $q^2$ bins are collected
in Table \ref{table: binned distributions of B to K nu nu}.
Our results for the integrated branching fraction
$\Delta {\cal BR}(0, (m_B-m_K)^2)= \left (6.02^{+1.68}_{-1.76} \right )  \times 10^{-6}$
are larger than the previous calculations \cite{Bartsch:2009qp}, where the authors employed
the rather small numbers of $f_{B K}^{+}(q^2=0)=0.304 \pm 0.042$, but they are still far below the
experimental upper bound from  BaBar Collaboration \cite{Lees:2013kla}.
Given the sizeable uncertainties for the predicted partial branching fractions of  $B \to K \nu \nu$,
we suggest to consider the ratio of the partial branching fractions for $B \to K \nu \nu$ and $B \to \pi \mu \nu_{\mu}$
\begin{eqnarray}
R_{K \pi}(q_1^2, q_2^2) = \frac{\int_{q_1^2}^{q_2^2} \, dq^2 \,\,
{d \, \Gamma(B \to K \nu \nu) / d q^2}}
{\int_{q_1^2}^{q_2^2} \, dq^2 \,\,  {d \, \Gamma(B \to \pi \mu \nu_{\mu}) /d q^2}}\,,
\end{eqnarray}
where the theory uncertainties due to the model-dependence of the $B$-meson LCDAs
are expected to be reduced significantly.
Our predictions for this ratio collected in Table \ref{table: binned distributions of B to K nu nu}
imply that the theory precision of $R_{K \pi}$ is approximately improved by
a factor of three, when compared with that of the partial branching fraction of $B \to K \nu \nu$.

\section{Summary}
\label{sect: summary}

In this paper we have computed the SU(3)-flavour symmetry breaking effects between
$B \to \pi$ and $B \to K$ form factors at large recoil from the LCSR with $B$-meson LCDAs.
It has been explicitly shown that the strange-quark-mass induced corrections are not suppressed
by $\Lambda/m_b$ in the heavy quark expansion and they also preserve the large-recoil symmetry
relations of  $B \to P$ form factors. We further evaluated the higher-twist corrections to the
semileptonic $B$-meson decay form factors from both the two-particle and three-particle $B$-meson LCDAs
with a complete parametrization of the corresponding light-cone matrix elements.
In particular, we constructed an alternative model for the three-particle twist-five
and twist-six $B$-meson LCDAs employing the method of QCD sum rules.
The asymptotic behaviours  of these higher-twist LCDAs in HQET at small quark and gluon momenta
from the resulting local duality model are consistent with that determined from the conformal
spins of the relevant fields. It is interesting to observe that the two-particle higher-twist corrections
from the twist-five LCDA  $\hat{g}_B^{-}(\omega, \mu)$ satisfy the symmetry relations of the soft form factors,
while the three-particle higher-twist contributions violate such relations already at tree level.

Inspecting the obtained sum rules for $B \to \pi, K$ form factors numerically, we observed that
the dominant higher-twist corrections come from the two-particle $B$-meson LCDA effects instead of
the three-particle contributions. Our predictions for the SU(3)-flavour symmetry violations are in nice
agreement with that obtained from the recent calculations with the light-meson LCSR approach.
Applying the $z$-series parametrization, the improved LCSR results of $B \to \pi, K$ form factors were extrapolated
to the whole kinematical region and compared with the Lattice QCD determinations.
Having in our hands the theory predictions for these form factors,
we computed the quantity $\Delta \zeta_{\mu}(0, 12 \, {\rm GeV^2})$
for the semileptonic $B \to \pi \mu \nu_{\mu}$ decay in  (\ref{zeta integral}),
from which the CKM matrix element $|V_{ub}|=\big( 3.23\,{}^{+0.66}_{-0.48} \big |_{\rm th.}\,{}^{+0.11}_{-0.11} \big |_{\rm exp.}  \big )
\times 10^{-3}$ was determined at the accuracy of ${\cal O} \, (20 \, \%)$.
The most significant theory uncertainty was identified to be generated by the variations of the inverse moment $\lambda_B(\mu_0)$.
Employing our results for the vector $B \to K$ form factor, we proceeded to compute the normalized differential $q^2$ distributions
of the rare exclusive $B \to K \nu \nu$ decays, which are expected to be well measured
(approximately $9 \, \%$  accuracy) at SuperKEKB  with the design luminosity forty times larger than that of KEKB.
In order to reduce the theory uncertainties, we constructed precision observables defined by the ratio of
 the partial branching fractions of $B \to K \nu \nu$ and $B \to \pi \ell \nu$.

Further improvements of the theory predictions for the heavy-to-light $B$-meson form factors
can be made in distinct directions. First, it would be interesting to improve the considered models
for the higher-twist $B$-meson LCDAs by taking into account the large-momentum behaviours from
perturbative QCD analysis. To this end, the classical EOM relations between the leading-twist and higher-twist
LCDAs  displayed in (\ref{the first EOM})-(\ref{the fourth EOM}) also need to be extended to the one-loop level.
Second, computing perturbative corrections to the higher-twist contributions in $B \to \pi, K$ form factors
is of both technical and conceptual importance for understanding  factorization properties of the exclusive
semileptonic $B$-meson decays. The one-loop evolution equations of the higher-twist $B$-meson LCDAs at
twist-six accuracy will be essential to such analysis. Third, improving the unitary bounds for the $z$-series
parametrizations of $B \to \pi, K$ form factors will be helpful to  constrain the momentum-transfer dependence
of these form factors (see \cite{Bigi:2016mdz} for further discussions on $B \to D$ form factors).

\subsection*{Acknowledgements}

C.D.L is supported in part by the National Natural Science Foundation
of China (NSFC) with Grant No. 11521505 and 11621131001.
The work of Y.L.S is supported by Natural Science Foundation of Shandong Province,
China under Grant No. ZR2015AQ006.
Y.M.W acknowledges support from the National Youth Thousand Talents Program,
the Youth Hundred Academic Leaders Program of Nankai University, and the NSFC with
Grant No. 11675082 and 11735010.


\appendix



\end{document}